\documentclass{article}
\usepackage[pdftex]{hyperref}
\usepackage[utf8]{inputenc}

\usepackage{changes}
\usepackage{xcolor,graphicx}
\usepackage{amsmath}
\usepackage[version=4]{mhchem}
\usepackage{siunitx}
\usepackage{longtable,tabularx}
\usepackage{physics}
\usepackage{amssymb}
\usepackage{physics}
\usepackage{graphicx}
\usepackage{mathrsfs}
\usepackage{color}
\usepackage[retainorgcmds]{IEEEtrantools}
\usepackage{newtxmath}
\usepackage{tikz}
\usepackage{standalone}
\usepackage{authblk}
\usepackage{natbib}
\setcitestyle{square,numbers,comma}
\usepackage[margin=1in]{geometry}

\newcommand{\Red}[1]{{\color{black}#1}}

\newcommand{\ii}{\mathrm{i}}
\newcommand{\ee}{\mathrm{e}}

\title{Revealing Structure and Symmetry of Nonlinearity in Natural and Engineering Flows}

\author[1,3]{Brandon Yeung}
\author[2,3]{Tianyi Chu}
\author[1,*]{Oliver T. Schmidt}
\affil[1]{Department of Mechanical and Aerospace Engineering, Jacobs School of Engineering, UCSD, 9500 Gilman Drive, La Jolla, CA 92093-0411, USA}
\affil[2]{School of Computational Science $\&$ Engineering, Georgia Institute of Technology, 756 West Peachtree Street NW, Atlanta, GA 30332-4017, USA}
\affil[3]{BY and TC are equal contributors to this work and designated as co-first authors.}
\affil[*]{oschmidt@ucsd.edu}
\date{}

\begin{document}

\maketitle

\begin{abstract}

Energy transfer across scales is fundamental in fluid dynamics, linking large-scale flow motions to small-scale turbulent structures in engineering and natural environments. Triadic interactions among three wave components form complex networks across scales, challenging understanding and model reduction. We introduce Triadic Orthogonal Decomposition (TOD), a method that identifies coherent flow structures optimally capturing spectral momentum transfer, quantifies their coupling and energy exchange in an energy budget bispectrum, and reveals the regions where they interact. TOD distinguishes three components—a momentum recipient, donor, and catalyst—and recovers laws governing pairwise, six-triad, and global triad conservation. Applied to unsteady cylinder wake and wind turbine wake data, TOD reveals networks of triadic interactions with forward and backward energy transfer across frequencies and scales.


\end{abstract}





\tableofcontents

\section{Introduction}\label{sec:intro}





Tones in spectra are a common feature of both natural and engineering flows, providing valuable insights into the extrinsic and intrinsic, as well as linear and nonlinear, dynamics of fluid systems. These tones can generally be classified into two categories: fundamental and nonlinear.

Fundamental tones either arise from external influences or originate from intrinsic flow instabilities. External frequencies are prevalent in many technical applications, such as the blade-passing frequency in rotating machinery like turbines, fans, and propellers. In geophysical and atmospheric flows, tonal peaks corresponding to semi-diurnal, diurnal, and annual frequencies are driven by Earth's rotation, axial tilt, and orbital dynamics relative to the Sun and Moon. Hydrodynamic instabilities, such as Kelvin-Helmholtz waves in shear layers, represent the second source of fundamental tones. These tones can often be predicted by hydrodynamic stability theory as discrete eigenvalues of the linearized dynamics. However, even linearly stable flows can exhibit tones when forced near an eigenfrequency, a phenomenon known as linear resonance.

Nonlinear tones, on the other hand, arise from the interactions of fundamental or other nonlinear tones. These originate from the quadratic nonlinearity of the convective term in the Navier-Stokes equations, which generates triadic sum- and difference-frequency interactions. These resonant triadic interactions produce peaks at sub-, super-, and ultra-harmonic frequencies. It is important to distinguish these from non-resonant triadic interactions, which only redistribute energy without producing distinct tones, as seen in the broadband spectrum of isotropic, homogeneous turbulence. While fundamental tones are often straightforward to identify by matching frequencies to known external sources or predictions from linear theory, nonlinear tones are typically identified by matching non-fundamental peaks in power spectra to integer multiples or sums and differences of fundamental frequencies. Since the power spectrum defines energy as the squared magnitude of the signal, it is inherently phase-blind. Thus, peaks at sub- and super-harmonic frequencies are a necessary, but not sufficient, condition for confirming nonlinearity. The occurrence of ultra-harmonic tones from triadic interactions between different spatial symmetry components adds further complexity to interpreting power spectra alone.


A well-established tool for identifying triadic interactions in pointwise signals is the bispectrum, along with its normalized counterpart, the bicoherence. The bispectrum is a third-order statistic that correlates two frequency components with their sum, making it useful for detecting triadic frequency interactions that satisfy the zero-sum condition by analyzing the phase correlation among the three involved frequencies (see \citet{brillinger1965introduction} for an early review). Building on this, \citet{schmidt2020bispectral} proposed bispectral mode decomposition (BMD), which uses an integral measure of bispectral density to identify coherent flow structures associated with triadic interactions. BMD achieves this by maximizing the block-to-block correlation between the local product of two frequency components and their sum, determining the numerical radius of a spectral estimate of the bispectral density matrix. While BMD identifies a single mode pair representing the dominant triad at each point in the bispectrum plane, the method introduced in this work provides complete bases of convective modes and recipient modes, along with a modal energy budget that quantifies the scale-to-scale energy transfer.

While the nonlinear term in the momentum equations can transfer and redistribute kinetic energy, it can neither produce nor remove net energy. The conservation of scale-to-scale kinetic energy transfer has been known at least as far back as the early text of \citet{Batchelor1953Cambridge}, and was later expounded on by \citet{kraichnan1959jfm} in the wavenumber domain. A systematic analysis of spectral energy conservation and transfer was also carried out by \citet{smyth1992pof}, which is the foundation for this work. The properties of inter-triad energy conservation are also treated in standard texts, including \citet{Lesieur1990Kluwer}, \citet{schmidhenningson2001springer}, and others. These properties are directly analogous in the frequency domain. Recent applications of kinetic energy transfer analysis and inter-triad conservation are found in the works of \citet{Barthel2022thesis} and \citet{FreemanEtAl2024JFM}. The transport of kinetic energy in fluid flows is governed by the dynamical equations for the mean kinetic energy (MKE) and turbulent kinetic energy (TKE). These equations may be partitioned into terms that can be interpreted as distinct mechanisms for energy flow, see e.g. \citet{pope2000cambridge}. They include: production, dissipation, advection, and the transfer of energy through pressure, viscous diffusion, and nonlinear interactions. Depending on the flow, each term represents either a contribution of energy to, or a removal of energy from, the overall kinetic energy budget. In Fourier space, the spectral kinetic energy equation describes the transfer of energy across scales, from frequency to frequency, and wavenumber to wavenumber. One of the pioneering examples of spectral energy budget analysis of inhomogeneous flows is the work by \citet{lumley1964pof}, which investigated the transfer of energy between wavenumbers in wall-bounded turbulence, and proposed the idea of an inverse energy cascade in which energy is transported from smaller to larger scales.
\Red{
\citet{smyth1992pof} later proposed a formalism that enables a systematic examination of the directivity of energy transfer from a donor scale to a recipient scale. Similar observations about energy directivity were also made by \citet{Batchelor1953Cambridge}. More recently, the donor-recipient interpretation has been used to study energy transfer in turbulence by \citet{DomaradzkiRogallo1990POF}, \citet{WebberEtAl2002IJNMF}, \citet{Alexakis2005PRL}, \citet{deWitEtAl2022JFM}, \citet{DingEtAl2024arxiv}, as well as in magnetohydrodynamic (MHD) turbulence by \citet{DarEtAl2001PhysD}, \citet{AlexakisEtAl2005PRE}, among others.}

Whereas early studies of energy budgets tended to emphasize energy transfer as a local (pointwise) phenomenon, \citet{RempferFasel1994jfm} demonstrated that it is possible to incorporate the concept of coherent structures into a framework of modal energy budget analysis. A comprehensive summary of the methodology and application to a complex bluff-body flow can be found in \citet{hosseini2016modal}. In these works, coherent structures were educed via space-only POD and employed as the basis for a Galerkin reduced-order model (ROM) of the global energy dynamics. The model thus emphasizes the transfer of energy mediated by interactions between structures. 
\Red{
Instead of using POD modes, \citet{jin2021energy} used optimal resolvent modes to compute the convective terms and energy transfers.
While these modal bases optimally represent the observable dynamics or the linear input-output system, TOD modes are specifically designed to optimally account for triad interactions.
}
This feature makes the latter ideally suited to the analysis of modal energy budgets in the frequency domain. Among the terms in the MKE and TKE budget equations, production, advection, and nonlinear transfer are triadic in nature. We show that these triadic energy budgets can be directly recovered from the convective modes and recipient modes of the proposed decomposition.


The remainder of this paper is organized as follows. First, \S \ref{sec:ortho_decomp} introduces the governing equations for linear momentum and kinetic energy in the frequency domain, and the new orthogonal decomposition in \S\S \ref{subsec:eqns} and \ref{sec:method}, respectively. Thereafter, \S \ref{sec:modalEnergyFlow} covers the modal energy flow analysis, starting with the theory of the spectral energy budget in \S \ref{subsec:budget} and inter-triad energy conservation in \S \ref{subsec:conservation}, before applying these concepts to modal decomposition in \S \ref{subsec:directivity}. Two applications of the decomposition are demonstrated in \S \ref{sec:applications}: the cylinder wake at $\Re=100$ as a canonical example of a nonlinear laminar flow in \S \ref{sec:cylinder}, and PIV data of a wind turbine wake as an example of a turbulent engineering flow in \ref{sec:windturbine} in \S \ref{sec:cylinder}. Finally, \S \ref{sec:discussion} discusses our findings and summarizes the method.

\section{Triadic Orthogonal Decomposition}\label{sec:ortho_decomp}

\subsection{Spectral Momentum and Kinetic Energy Equations}\label{subsec:eqns}
The motion of an incompressible Newtonian fluid is governed by the momentum equations,
\begin{align}\label{mom_eqn}
    \pdv{\vb*u}{t} = -(\vb*u\vdot\grad)\vb*u - \grad p + \frac{1}{\mathrm{Re}}\laplacian\vb*u,
\end{align}
where $\vb*u = [u\;v\;w]^T$ the velocity vector field. Equations (\ref{mom_eqn}) have been nondimensionalized by the velocity scale, $U_\infty$, the length scale, $L$, and are parameterized by the Reynolds number, $\mathrm{Re}=\frac{U_\infty L}{\nu}$, where $\nu$ is the kinematic viscosity. 
Assuming a time-periodic flow, the flow variables can be expanded as the Fourier series,
\begin{align}
    \vb*u(\vb*x,t) = \sum_{n=-\infty}^\infty \hat{\vb*u}_n(\vb*x)\ee^{\ii 2\pi f_n t} \qqtext{and} p(\vb*x,t) = \sum_{n=-\infty}^\infty \hat{p}_n(\vb*x)\ee^{\ii 2\pi f_n t}.
\end{align}
For each frequency, $f_n$, we obtain the frequency-domain representation of equation~\eqref{mom_eqn},
\begin{align}\label{spec_mom_eqn}
    \ii 2\pi f_n \hat{\vb*u}_n &= -\qty(\widehat{(\vb*u\vdot\grad)\vb*u})_n - \grad \hat{p}_n + \frac{1}{Re}\laplacian\hat{\vb*u}_n = -\sum_{l=-\infty}^\infty \underbrace{(\hat{\vb*u}_{n-l}\vdot\grad)\hat{\vb*u}_l}_{-\vb*{\hat{c}}_{l\rightarrow n}} - \grad \hat{p}_n + \frac{1}{Re}\laplacian\hat{\vb*u}_n ,
\end{align}
where in the second step we have invoked the convolution theorem. This frequency-domain, or spectral, momentum equation lies at the core of our modal momentum transfer analysis. The special role and implications of the components in the convolution sum, $\vb*{\hat{c}}_{l\rightarrow n}$, are discussed in \S \ref{sec:recdon}.

For the often-used notation of triads, $(k,l,n)$ with $k=n-l$, any two of the frequency components determine the remaining one.
For simplicity, we use the pairwise notation $(l,n)$ to denote the triplet $(n-l,l,n)$ in the following. For the case of $n=0$,
equation~\eqref{spec_mom_eqn} simplifies to
\begin{align}\label{eq:RANS}
    0 = -(\bar{\vb*u}\vdot\grad)\bar{\vb*u} -\sum_{l\neq0} \underbrace{(\hat{\vb*u}_{-l}\vdot\grad)\hat{\vb*u}_l}_{-\vb*{\hat{c}}_{l\rightarrow 0}} - \grad \bar{p} + \frac{1}{Re}\laplacian\bar{\vb*u} = -(\bar{\vb*u}\vdot\grad)\bar{\vb*u} - \overline{(\vb*u'\vdot\grad)\vb*u'} - \grad \bar{p} + \frac{1}{Re}\laplacian\bar{\vb*u},
\end{align}
which are the Reynolds-averaged Navier-Stokes (RANS) equations, with $\vb*u'\equiv\vb*u-\bar{\vb*u}$. The definition of $\vb*u'$ implies that $\hat{\vb*u}_l=\widehat{\vb*u'_l}$ for $l\neq0$, which leads to the second equality in equation~\eqref{eq:RANS}. Each term in the RANS equations contributes to the deformation of the mean flow, $\bar{\vb*u}$. The term $\vb*{\hat{c}}_{l\rightarrow 0}$, in particular, is responsible for momentum transfer via the Reynolds stress, $\overline{(\vb*u'\vdot\grad)\vb*u'}$.

Upon left-multiplying with $\hat{\vb*u}_n^{\mathrm{H}}$, equation~\eqref{spec_mom_eqn} can be rewritten as the spectral kinetic energy equation,
\begin{align}\label{spec_mom_eqn_mult}
\ii 2\pi f_n\hat{\vb*u}_n^{\mathrm{H}}\hat{\vb*u}_n = -\hat{\vb*u}_n^{\mathrm{H}}\sum_{l=-\infty}^\infty\underbrace{(\hat{\vb*u}_{n-l}\vdot\grad)\hat{\vb*u}_l}_{-\vb*{\hat{c}}_{l\rightarrow n}} - \div(\hat{p}_n\hat{\vb*u}_n^{\mathrm{H}}) + \frac{1}{Re}\hat{\vb*u}_n^{\mathrm{H}}\laplacian\hat{\vb*u}_n,
\end{align}
where we used expression \ref{eq:c} for the convective term. The spectral kinetic energy equation governs the distribution of the kinetic energy, $\hat{E}_n=\frac{1}{2}\hat{\vb*u}_n^{\mathrm{H}}\hat{\vb*u}_n$, and forms the basis for the modal energy flow analysis in \S \ref{sec:modalEnergyFlow}. \Red{A remarkable property of equation (\ref{spec_mom_eqn_mult}) regarding the convective term, which is not used here but discussed in detail by \citet{FreemanEtAl2024JFM}, is that under certain conditions, the imaginary part of the term can be used to recover the integral power of the velocity, and the real part the integral dissipation.}

A wide variety of open flows exhibit strong convective instability.
Examples encompass natural and technical flows, such as bluff-body wakes and boundary layers, as well as exogenously forced flows like plasma-actuated jets.
In these flows, the convective term $\vb*{\hat{c}}_{l\rightarrow n}$ becomes dominant in both spectral momentum and kinetic energy equations. In the following, we outline how the spatial structures involved in spectral momentum transfer can be leveraged to identify and understand triadic flow interactions.

\subsection{Recipient-Donor Framework}\label{sec:recdon}
The recipient-donor framework of the momentum equations is a key concept that facilitates the calculation of both the direction and magnitude of modal energy transfer. In equation~\eqref{spec_mom_eqn}, we introduced the notation
\begin{equation}\label{eq:c}
    \vb*{\hat{c}}_{l\rightarrow n} = -(\hat{\vb*u}_{n-l}\vdot\grad)\hat{\vb*u}_l
\end{equation}
to represent the component in the convolution sum associated with the frequency pair $(l,n)$, or the corresponding triplet $(n-l,l,n)$. This notation implicitly denotes the direction of momentum transfer, $l\rightarrow n$, from frequency component $\hat{\vb*u}_l$ to $\hat{\vb*u}_n$. As equation~\eqref{spec_mom_eqn} governs the velocity component at frequency $f_n$, we denote the component $\hat{\vb*u}_n$ as the \emph{recipient} of linear momentum (though contributions can be negative, resulting in a net loss of momentum). We first present an intuitive argument based on the roles of $\hat{\vb*u}_{n-l}$ and $\hat{\vb*u}_l$ in $\vb*{\hat{c}}_{l\rightarrow n}$ in the familiar derivation of the momentum equations, which indicates that the velocity component $\hat{\vb*u}_{n-l}$ advects $\hat{\vb*u}_l$ in the direction of, and with a magnitude proportional to, its gradient field, $\nabla \hat{\vb*u}_l$. Thus $\hat{\vb*u}_l$ serves as a \emph{donor} of linear momentum to $\hat{\vb*u}_n$, while $\hat{\vb*u}_{n-l}$ acts as the passive advector. To avoid confusion with the \emph{convective field}, $\vb*{\hat{c}}_{l\rightarrow n}$, which ultimately contributes  to $\hat{\vb*u}_n$ and hence frequency $f_{n}$ we refer to $\hat{\vb*u}_{n-l}$ as the \emph{catalyst field}, and $f_{n-l}$ as the \emph{catalyst frequency}.

From the derivation and final form of equation (\ref{spec_mom_eqn_mult}), which governs the spectral kinetic energy $\hat{E}_n=\frac{1}{2}\hat{\vb*u}_n^{\mathrm{H}}\hat{\vb*u}_n$, it is evident that spectral energy transfer to frequency $f_{n}$ results directly from momentum transfer by the \emph{convective field}, $\vb*{\hat{c}}_{l\rightarrow n}$. The donor-recipient interpretation from $f_{l}$ to $f_{n}$ therefore directly translates to the transfer of energy, and can, in fact, be derived from the inter-triad energy conservation discussed in \S \ref{subsec:conservation}. Specifically, under certain conditions, the energy transfer by triad $(n-l,l,n)$ is exactly equal and opposite to that by the triad $(l-n,n,l)$ and hence independent of the catalyst component. This second, rigorous motivation of the recipient-donor interpretation is elaborated in \S \ref{subsec:directivity}.

\subsection{Modal Decomposition}\label{sec:method}



The recipient-donor interpretation of the spectral momentum and kinetic energy equations, as summarized in \S\ref{sec:recdon}, establishes the relationship where $\hat{\vb*u}_n$ acts as the \emph{recipient} of momentum or energy through the action of the \emph{convective field}, $\vb*{\hat{c}}_{l\rightarrow n}$. The objective of decomposition is to identify the flow structures that, \emph{on average}, contribute most to the momentum transfer. This objective can be framed as an optimization problem using the calculus of variations, where a pair of jointly optimal modal bases for a given triad, $(l,n)$, are computed such that the convective term, $\vb*{\hat{c}}_{l\to n}$, and the recipient, $\vb*{\hat{u}}_{n}$, are optimally represented in terms of their covariance.
The kinetic energy of the convective term and the recipient
are measured in the norm $\|\cdot\|$, induced by the inner product 
\begin{equation}
\expval*{\vb*u_1(\vb*{x}),\vb*u_2(\vb*{x})} =  \int_{\Omega} \vb*u_2^{\mathrm{H}}(\vb*{x}) \vb*u_1(\vb*{x}) \, \dd \vb*{x}.
\end{equation}
The optimal convective-recipient covariance for each {donor-recipient pair, $\sigma(l,n)\in \mathbb{R}_{\geq0}$}, is given by the optimization problem 
\begin{subequations}\label{eq:optmCov}
\begin{align}
    \sigma(l,n) 
    & = \max \frac{\mathrm{E}\qty{ \expval*{ \vb*{\hat{c}}_{l\to n}(\vb*{x}),\vb*{\hat{\psi}}_{l\to n}(\vb*{x})} \expval*{\vb*{\hat{u}}_{n}(\vb*{x}'),\vb*{\hat{\phi}}_{n}(\vb*{x}')}^* }}{ \norm*{  \vb*{\hat{\psi}}_{l\to n}(\vb*{x})} \norm*{\vb*{\hat{\phi}}_n(\vb*{x}')} }\\
    & = \max \frac{\mathrm{E}\qty{ \left(\int_{\Omega}   \vb*{\hat{\psi}}_{l\to n}^{\mathrm{H}}(\vb*{x}) \vb*{\hat{c}}_{l\to n}(\vb*{x}) \, \dd \vb*{x} \right)\left(  \int_{\Omega} \vb*{\hat{u}}_{n}^{\mathrm{H}}(\vb*{x}')  \vb*{\hat{\phi}}_{n}(\vb*{x}') \, \dd \vb*{x}'  \right) }}{ \norm*{  \vb*{\hat{\psi}}_{l\to n}(\vb*{x})}  \norm*{\vb*{\hat{\phi}}_n(\vb*{x}')}   } \\
    & = \max \frac{\mathrm{E}\qty{ \iint_{\Omega}   \vb*{\hat{\psi}}_{l\to n}^{\mathrm{H}}(\vb*{x}) \left( \vb*{\hat{c}}_{l\to n}(\vb*{x}) \vb*{\hat{u}}^{\mathrm{H}}_{n}(\vb*{x}')  \right)   \vb*{\hat{\phi}}_{n}(\vb*{x}') \, \dd \vb*{x} \, \dd \vb*{x}'  }}{ \norm*{  \vb*{\hat{\psi}}_{l\to n}(\vb*{x})}  \norm*{\vb*{\hat{\phi}}_n(\vb*{x}')}   } \\
     & = \max \frac{ \iint_{\Omega}   \vb*{\hat{\psi}}_{l\to n}^{\mathrm{H}}(\vb*{x}) \vb*{S}(\vb*x,\vb*x';n,l)  \vb*{\hat{\phi}}_{n}(\vb*{x}') \, \dd \vb*{x} \, \dd \vb*{x}'  }{ \norm*{  \vb*{\hat{\psi}}_{l\to n}(\vb*{x})}  \norm*{\vb*{\hat{\phi}}_n(\vb*{x}')}   },
\end{align}
 \end{subequations} 
where
\begin{align}
    \vb*{S}(\vb*x,\vb*x';l,n) =  \mathrm{E}\qty{\vb*{\hat{c}}_{l\to n}(\vb*x)\vb*{\hat{u}}^{\mathrm{H}}_{n}(\vb*x')}
\end{align}
represents the two-point cross-bispectral covariance tensor. This tensor is a Fredholm kernel, and the modes that jointly maximize $\sigma$ can be obtained via the
singular value expansion (SVE) of $\vb*{S}$ \citep{schmidt1907theorie}, that is 
\begin{align}\label{Sfq_SVD}
    \vb*{S}(\vb*x,\vb*x';l,n) = \sum_{j=1}^{\infty} \sigma_j {  \vb*{\hat{\psi}}}_{l\to n,j}(\vb*{x})\vb*{\hat{\phi}}^{\mathrm{H}}_{n,j}(\vb*{x}').
\end{align}
The reality and non-negativeness of $\sigma$ are guaranteed by the SVE.
The convective mode, $   \vb*{\hat{\psi}}_{l\to n,j}$, and recipient mode, $\vb*{\hat{\phi}}_{n,j}$, are ordered by their associated singular values, ${\sigma_1}\geq {\sigma_2}\geq \cdots \geq 0$. They are also orthogonal in their respective inner products and have unit energy, that is, $\left<  \vb*{\hat{\psi}}_{l\to n,i},  \vb*{\hat{\psi}}_{l\to n,j}\right>  =  \left<\vb*{\hat{\phi}}_{n,i},\vb*{\hat{\phi}}_{n,j}\right>   =\delta_{ij}$. In practice, that is for discrete data, equation (\ref{Sfq_SVD}) is solved by the singular value decomposition.
In the special case where the convective term is identical to the recipient, $\vb*{\hat{c}}_{l\to n}= \vb*{\hat{u}}_{n}$, equations (\ref{eq:optmCov}-\ref{Sfq_SVD}) reduce to the proper orthogonal decomposition framework by \citet{lumley1967structure,lumley1970stochastic}.
By defining the expansion coefficients as the projections of the convective term and the recipient onto their respective modal bases (suppressing their dependence on $l$ and $n$ for clarity),
 \begin{align}\label{triadic_lowD}
     a_j = \left<\vb*{\hat{u}}_{n},\vb*{\hat{\phi}}_{n,j}\right>  \quad \text{and} \quad 
    b_j = \left<  \vb*{\hat{c}}_{l\to n},\vb*{\hat{\psi}}_{l\to n,j}\right> ,
 \end{align}
 we obtain the jointly optimal, ranked decomposition,
\begin{align}\label{eq:modalExpansion}
    \vb*{\hat{u}}_{n} = \sum_{j= 1}^{\infty} a_j \vb*{\hat{\phi}}_{n,j} \quad \text{and} \quad \vb*{\hat{c}}_{l\to n}= \sum_{j= 1}^{\infty} b_j \vb*{\hat{\psi}}_{l\to n,j}.
\end{align}
The expansion coefficients, $a$ and $b$, satisfy
\begin{equation}\label{eq:coeffsUncorrelated}
    \mathrm{E}\{ b_j a_k^{*} \} = \sigma_j \delta_{jk}.
\end{equation}
In other words, the coefficients are uncorrelated with one another. \Red{Since the coefficients are complex quantities that retain phase information, the covariance between $a_j$ and $b_j$, i.e., the singular value, additionally measures the degree of phase coupling between the projections of the recipient, $\vb*{\hat{u}}_{n}$, and the convective term, $\vb*{\hat{c}}_{l\to n}$.} 

To obtain convergent estimates of the bispectral densities, we adapt Welch's \citep{welch1967use} method in the decomposition framework to construct an ensemble of realizations of the temporal Fourier transform of the data from a single time series consisting of $N_t$ snapshots based on the ergodicity hypothesis for statistically stationary flows. First, the data is segmented into $N_\mathrm{blk}$, potentially overlapping by $N_\mathrm{ovlp}$ snapshots, blocks containing $N_f$ consecutive snapshots separated by time step $\Delta t$. Each snapshot is represented by a column vector of length $N$, corresponding to the number of spatial degrees of freedom times the number of velocity components, i.e., the dimension of the data. The $m$th block is defined as
\begin{align}
     \mathbf{{U}}^{(m)} &=   \left[
\begin{matrix}
  \mathbf{{u}}_{1}^{(m)}& \mathbf{{u}}_{2}^{(m)} &\cdots & \mathbf{{u}}_{N_f}^{(m)}
\end{matrix}\right]  \in \mathbb{C}^{N\times N_f},
\end{align}
and its temporal, i.e., row-wise, DFT is denoted by
\begin{align}
\mathbf{\hat{U}}^{(m)} &=   \left[
\begin{matrix}
  \mathbf{\hat{u}}_{1}^{(m)}& \mathbf{\hat{u}}_{2}^{(m)} &\cdots & \mathbf{\hat{u}}_{N_f}^{(m)}
\end{matrix}\right], 
\end{align}
respectively. This is the same spectral estimation technique commonly used in spectral proper orthogonal decomposition (SPOD), and readers are referred to \citet{towne2018spectral,schmidt2020guide} for more details and best practices. Next, the data matrices for the recipient and convective terms,  
\begin{subequations}
    \begin{align}
        \mathbf{\hat{U}}_{n} &=  
        \left[
\begin{matrix}
  \mathbf{\hat{u}}_{n}^{(1)}& \mathbf{\hat{u}}_{n}^{(2)} &\cdots & \mathbf{\hat{u}}_{n}^{(N_\mathrm{blk})}
\end{matrix}\right], \\
        \mathbf{\hat{C}}_{l\to n} &= 
        \left[
\begin{matrix}
  \vb{\hat{c}}_{l\to n}^{(1)} & \vb{\hat{c}}_{l\to n}^{(2)} & \cdots & \vb{\hat{c}}_{l\to n}^{(N_\mathrm{blk})}
\end{matrix}\right] = 
-\left[
\begin{matrix}
(\hat{\vb u}_{n-l}^{(1)}\vdot\grad)\hat{\vb u}_l^{(1)} & (\hat{\vb u}_{n-l}^{(2)}\vdot\grad)\hat{\vb u}_l^{(2)} & \cdots &(\hat{\vb u}_{n-l}^{(N_\mathrm{blk})}\vdot\grad)\hat{\vb u}_l^{(N_\mathrm{blk})}
\end{matrix}\right],   \end{align}
\end{subequations}
are assembled from the columns of $\mathbf{\hat{U}}^{(m)}$. Then, the two-point cross-bispectral covariance tensor at triad $(l,n)$ can be approximated as
\begin{align}
    \mathbf{S}_{l,n} = {\frac{1}{N_\mathrm{blk}}}\mathbf{\hat{C}}_{l\to n} \mathbf{\hat{U}}_{n}^{\mathrm{H}}.
\end{align}
Using this approximation, the infinite-dimensional SVE problem in equation (\ref{Sfq_SVD}) reduces to an $N\times N$ matrix SVD problem
\begin{align}
   \vb W^{1/2}\mathbf{S}_{l,n}\vb W^{1/2} &= \mathbf{\tilde{\Psi}}_{l\to n}\mathbf{\Sigma}_{l,n}\mathbf{\tilde{\Phi}}_{n}^{\mathrm{H}},
\end{align}
for the convective-recipient covariances, $\mathbf{\Sigma}_{l,n}=\text{diag}([\sigma_1, \sigma_2, \dots, \sigma_{N_\mathrm{blk}}])$, captured by each mode pair. The convective and recipient modes can be recovered as 
\begin{align}
  \mathbf{\hat{\Psi}}_{l\to n} =\vb W^{-1/2} \mathbf{\tilde{\Psi}}_{l\to n}, \qand 
   \mathbf{\hat{\Phi}}_{n} = \vb W^{-1/2}\mathbf{\tilde{\Phi}}_{n},
\end{align}
respectively, and $\mathbf{W}$ is a weight matrix that accounts for the quadrature weights. In practice, there is no need to directly solve this SVD problem as the matrix $\mathbf{S}$ is low-rank with a rank of $N_\mathrm{blk}\ll N$. Instead, we first perform the QR decomposition of the weighted convective term data matrix 
\begin{align}
    \frac{1}{N_\mathrm{blk}}{\vb W^{1/2}}\mathbf{\hat{C}}_{l\to n} =  \vb C_\perp  \vb C_\triangle,
\end{align}
where $\vb C_\perp\in \mathbb{C}^{N\times N_\mathrm{blk}}$ is orthonormal and $\vb C_\triangle\in \mathbb{C}^{N_\mathrm{blk}\times N_\mathrm{blk}}$ is upper triangular. 
We then compute the EVD of the Hermitian matrix $\vb C_\triangle \mathbf{\hat{U}}_{n}^{\mathrm{H}} {\vb W} \mathbf{\hat{U}}_{n} \vb C_\triangle^{\mathrm{H}} \in \mathbb{C}^{N_\mathrm{blk}\times N_\mathrm{blk}}$ such that
    \begin{align}
       \left(\vb C_\triangle \mathbf{\hat{U}}_{n}^{\mathrm{H}} {\vb W} \mathbf{\hat{U}}_{n} \vb C_\triangle^{\mathrm{H}}\right){\mathbf{{\Psi}}_{l\to n}}  = {\mathbf{{\Psi}}_{l\to n}}\mathbf{\Lambda}_{l\to n}.
    \end{align}
The convective modes, recipient modes, and associated singular values can then be recovered as 
\begin{align}
    \mathbf{\hat{\Psi}}_{l\to n}&=\mathbf{W} ^{-1/2} \vb C_\perp {\mathbf{{\Psi}}_{l\to n}}, \\
    \mathbf{\hat{\Phi}}_{n}&= \mathbf{\hat{U}}_{n} \vb C_\triangle^{\mathrm{H}} {\mathbf{{\Psi}}_{l\to n}} \mathbf{\Lambda}_{l\to n}^{-1/2},  \\
    \mathbf{\Sigma}_{l\to n} &= \mathbf{\Lambda}_{l\to n}^{1/2} , 
\end{align}
respectively (the implicit dependence of $\mathbf{\hat{\Phi}}_{n}$ on $l$ is suppressed for readability). Alternatively, they form the weighted SVD of the sample cross-bispectral covariance matrix,
\begin{align}
    \vb S_{l,n} =  \vb{\hat\Psi}_{l\to n} \vb{\Sigma}_{l\to n} \vb{\hat\Phi}_{n}^\mathrm{H}, \qq{with} \vb{\hat\Phi}_{n}^\mathrm{H} \vb W \vb{\hat\Phi}_{n} = \vb{\hat\Psi}_{l\to n}^\mathrm{H} \vb W \vb{\hat\Psi}_{l\to n} = \vb I.
\end{align}

The expansion coefficients associated with the recipient and convective modes are respectively obtained by orthogonal projection,
\begin{equation}
    \vb A_{l,n} = \vb{\hat\Phi}_{n}^\mathrm{H} \vb W \vb{\hat U}_{n} \qand \vb B_{l,n} = \vb{\hat\Psi}_{l\to n}^\mathrm{H} \vb W \vb{\hat C}_{l\to n},
\end{equation}
from which the data matrices may be reconstructed as
\begin{equation}
    \vb{\hat U}_{n}=\vb{\hat\Phi}_{n}\vb A_{l,n} \qand \vb{\hat C}_{l\to n}=\vb{\hat\Psi}_{l\to n}\vb B_{l,n}.
\end{equation}


\subsection{Relationship to other methods}\label{sec:other}
In the literature, several data-driven methods bear similarities to TOD. Here, we briefly outline these methods without delving into extensive detail. The optimization problem in equation~\eqref{eq:optmCov} can be interpreted as maximizing the covariance between the expansion coefficients, expressed as \( \sigma_j = \max \, \mathrm{E}\{ b_j a_j^{*} \} \), suggesting that TOD can be seen as a specialized form of maximum covariance analysis (MCA) \citep{von1999statistical}. MCA, along with the closely related canonical correlation analysis (CCA) \citep{hotelling1992relations}, identifies bases that are jointly optimal in multivariate statistical terms and is widely applied in data mining. For further details on CCA and its variations, see \citep{thompson1984canonical, hardoon2004canonical, andrew2013deep}.

TOD and BMD \citep{schmidt2020bispectral} can be compared as both are bispectral methods that directly consider third-order statistics. As discussed in the introduction (\S \ref{sec:intro}), BMD focuses on block-to-block correlations, whereas TOD decomposes spatial correlations and explicitly distinguishes between convective and recipient terms. Despite these differences, both methods yield qualitatively consistent results. A related approach is cross POD (CPOD; \citep{CavalieridaSilva2021PRF}), which, despite its name, maximizes block-to-block correlations and therefore aligns more closely with BMD. Mathematically, BMD addresses the numerical radius problem, while CPOD solves for the numerical abscissa. CPOD could also be adapted to the frequency domain for specialized triad identification.

The variational formulation of TOD in equation \eqref{eq:optmCov} suggests connections to classical POD and its various extensions. Balanced POD (BPOD; \citep{rowley2005model}) seeks a pair of bi-orthogonal modal bases that optimally represent the controllability and observability Gramians in a linear input-output system. TOD can also be related to frequency-domain POD variants: spectral POD (SPOD; \citep{towne2018spectral, schmidt2020guide}) and cyclostationary SPOD (CS-SPOD; \citep{HeidtColonius2024JFM}). While SPOD optimally captures power but assumes frequency independence, it has been recently applied to study modal energy transfer \citep{nekkanti2024nonlineardynamicsvortexpairing}. CS-SPOD extends SPOD by including correlations between frequency components that become triadically linked in the presence of a time-periodic mean flow. An additional, but partially operator-driven extension for analyzing triadic interactions is the resolvent-based extended SPOD (RESPOD; \citep{TowneEtAl2015AIAA, KarbanEtAl2022JFM, KarbanEtAl2024arXiv}), which finds modes representing the convective term that share the same expansion coefficients as the usual SPOD modes.

\section{Modal Energy Flow Analysis}\label{sec:modalEnergyFlow}
The nonlinear term in the Navier-Stokes equations is conservative (see, e.g., \citet{schmidhenningson2001springer}). Unlike the linear terms, which may inject energy into or siphon energy out of the system, the nonlinear term only redistributes energy. Collectively, the net effect of all nonlinear triadic interactions is thus to conserve energy. Energy is also conserved by individual groups of triads \citep{kraichnan1959jfm}. Based on pairwise triadic conservation, \citet{smyth1992pof} proposed a methodology by which the energy exchanged between two arbitrary frequencies can be systematically quantified. We demonstrate the utility of this method for TOD and identify and quantify energy propagation via mode interactions.

\subsection{Spectral Energy Budget}\label{subsec:budget}   

\begin{figure}[ht]
    \centering
    \includegraphics[width=\textwidth]{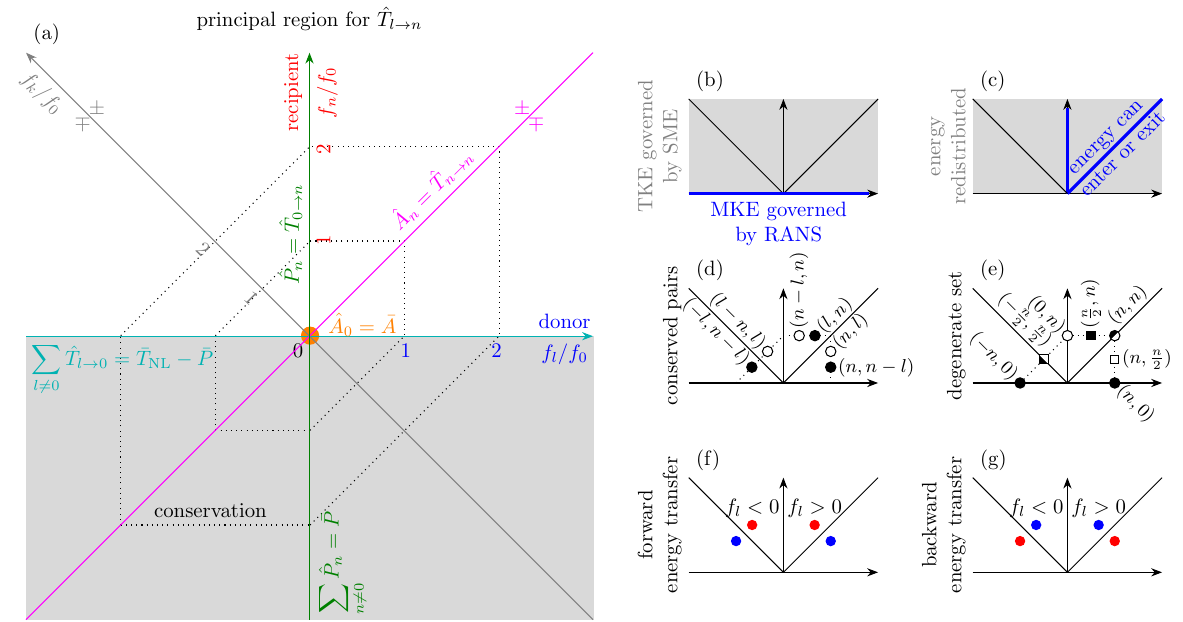}
    \caption{(a) Schematic of triadic energy transfer, $\hat{T}_{l\to n}$, in the bispectral plane with the donor frequency, $f_l$, as abscissa (blue) and the recipient frequency, $f_n$, as ordinate (red), both normalized by $f_0$, the fundamental frequency. Only the principal region has to be considered; the gray bottom-half plane contains redundant information. The catalyst frequency, $f_k=f_{n-l}$, is indicated as a gray line. $\hat{T}_{l\to n}$ is conserved on nested hexagons with dotted lines exemplifying this conservation for $f_l=f_n=f_0$ and $f_l=f_n=2f_0$. Further indicated are spectral TKE and MKE contributions from linear advection by the mean (magenta), production (green), mean self-advection (orange), and transfer-production difference (teal). The magenta $\pm$ symbols denote the property of pairwise conservation about $f_l=f_n$; similarly, the gray $\pm$ symbols denote conservation about $f_l=-f_n$. Panels (b--g) are interpretation aids.}
    \label{fig:principal_sketch}
\end{figure}

Taking the real part of equation~\eqref{spec_mom_eqn_mult} yields the conservation equation for the spectral kinetic energy,
\begin{align}\label{spec_k_eqn}
0 = \mathcal{R}\underbrace{\qty{-\hat{\vb*u}_n^{\mathrm{H}}\sum_{l=-\infty}^\infty(\hat{\vb*u}_{n-l}\vdot\grad)\hat{\vb*u}_{l}}}_{\hat{T}_n} + \mathcal{R}\underbrace{\qty{-\div(\hat{p}_n\hat{\vb*u}_n^{\mathrm{H}})}}_{\hat{F}_n} + \mathcal{R}\underbrace{\qty{\frac{1}{Re}\hat{\vb*u}_n^{\mathrm{H}}\laplacian\hat{\vb*u}_n}}_{\hat{D}_n}
\end{align}
which partitions the budget of the kinetic energy at frequency $f_n$, $\hat{E}_n=\frac{1}{2}\hat{\vb*u}_n^{\mathrm{H}}\hat{\vb*u}_n$, into contributions from inter-scale transfer, $\hat{T}_n$, pressure work, $\hat{F}_n$, and dissipation, $\hat{D}_n$. The transfer term may be expressed as the summation over individual triads,
\begin{align}
\hat{T}_n = \sum_{l=-\infty}^\infty \hat{T}_{l\to n},
\end{align}
where $\hat{T}_{l\to n} \equiv -\hat{\vb*u}_n^{\mathrm{H}}(\hat{\vb*u}_{n-l}\vdot\grad)\hat{\vb*u}_{l}$. The inter-scale transfer can be further partitioned into advection, $\hat{A}_n$, production, $\hat{P}_n$, and nonlinear transfer, $\hat{T}_{\mathrm{NL},n}$,
\begin{subequations}
\begin{align}
\hat{T}_n &= \hat{T}_{n\to n} + \hat{T}_{0\to n} + \sum_{l\ne0,n}\hat{T}_{l\to n} \\
&= \underbrace{-\hat{\vb*u}_n^{\mathrm{H}}(\bar{\vb*u}\vdot\grad)\hat{\vb*u}_n}_{\hat{A}_n} \underbrace{-\hat{\vb*u}_n^{\mathrm{H}}(\hat{\vb*u}_n\vdot\grad)\bar{\vb*u}}_{\hat{P}_n} \underbrace{-\hat{\vb*u}_n^{\mathrm{H}}\sum_{l\ne0,n}(\hat{\vb*u}_{n-l}\vdot\grad)\hat{\vb*u}_{l}}_{\hat{T}_{\mathrm{NL},n}}.
\end{align}
\end{subequations}
The terms $\hat{A}_n$ and $\hat{P}_n$ are triads in which one component is the mean flow, $\bar{\vb*u}$. They are, therefore, linear mechanisms with respect to the mean. The budget of $\hat{T}_n$ is summarized graphically in the bispectrum in figure \ref{fig:principal_sketch}(a), \Red{in which $\hat{A}_n$ and $\hat{P}_n$ are naturally recovered as special cases of $\hat{T}_{l\to n}$}. Along $l=n$, $\hat{T}_{l\to n}=\hat{A}_n$ and is indicated in magenta. Along $l=0$, $\hat{T}_{l\to n}=\hat{P}_n$ and is indicated in green. In the remainder of the bispectrum, quadratic nonlinearities are active.

For the special case of $n=0$, equation~\eqref{spec_k_eqn} simplifies to
\begin{subequations}\label{eq:mean_k_eqn_all}
\begin{align}\label{mean_k_eqn}
0 &= -\bar{\vb*u}^\mathrm{T}\left((\bar{\vb*u}\vdot\grad)\bar{\vb*u}+\sum_{l\neq 0}\mathcal{R}\underbrace{\qty{(\hat{\vb*u}_{-l}\vdot\grad)\hat{\vb*u}_{l}}}_{-\vb*{\hat{c}}_{l\rightarrow 0}}\right) -\div(\bar{p}\bar{\vb*u}) + \frac{1}{Re}\bar{\vb*u}^{\mathrm{T}}\laplacian\bar{\vb*u} \\
&=  \underbrace{-(\bar{\vb*u}\vdot\grad)\frac{1}{2}\bar{\vb*u}^\mathrm{T}\bar{\vb*u}}_{\bar A} \underbrace{-\div(\overline{\vb*u'(\vb*u')^\mathrm{T}}\bar{\vb*u})}_{\bar T_\mathrm{NL}} +\underbrace{\overline{\vb*u'(\vb*u')^\mathrm{T}}\boldsymbol{:}\grad\bar{\vb*u}}_{-\bar P} \underbrace{-\div(\bar{p}\bar{\vb*u})}_{\bar F} + \underbrace{\frac{1}{Re}\bar{\vb*u}^{\mathrm{T}}\laplacian\bar{\vb*u}}_{\bar D}, 
\end{align}
\end{subequations}
which is the evolution equation for the mean kinetic energy (MKE), $\bar E=\frac{1}{2}\bar{\vb*u}^\mathrm{T}\bar{\vb*u}$. The terms labeled $\bar A$, $\bar T_\mathrm{NL}$, $\bar P$, $\bar F$, and $\bar D$ are the contributions to the MKE from advection, transfer, production, pressure work, and dissipation, respectively. The mean production can alternatively be expressed as a summation of the spectral production,
\begin{equation}
\bar P = \sum_{n\neq0}\hat{P}_n = -\sum_{n\neq0}\hat{\vb*u}_n^{\mathrm{H}}(\hat{\vb*u}_n\vdot\grad)\bar{\vb*u} = -\overline{\vb*u'(\vb*u')^\mathrm{T}}\boldsymbol{:}\grad\bar{\vb*u},
\end{equation}
where we have invoked the general form of Parseval's Theorem to arrive at the last step. The MKE equation \eqref{eq:mean_k_eqn_all} is the counterpart to the RANS equations \eqref{eq:RANS}. The MKE budget can similarly be inferred from figure \ref{fig:principal_sketch}(a). Specifically, the mean transfer and production, $\bar T_\mathrm{NL}$ and $\bar P$, respectively, can be recovered from $\hat{T}_{l\to n}$ along the $f_n=0$ and $f_l=0$ axes. The mean advection,  $\bar A=\hat{T}_{0\to 0}$, is located at the origin. Together, equations~\eqref{eq:RANS} and \eqref{eq:mean_k_eqn_all} describe the deformation of the mean flow in terms of its momentum and energy, respectively. In particular, the convective term, $\vb*{\hat{c}}_{l\rightarrow 0}$, and its corresponding convective mode, $\hat{\vb*\psi}_{l\to 0}$, for all $l$ conspire to alter the mean flow by removing momentum and energy from the mean.

To aid in the interpretation of the bispectrum, each panel in figure \ref{fig:principal_sketch}(b--g) highlights a distinct concept. Figure \ref{fig:principal_sketch}(b) illustrates the region of the bispectrum plane governed by the spectral momentum equations (SME), $f_n\neq0$, and the region governed by the RANS equations, $f_n=0$. Figure \ref{fig:principal_sketch}(c) illustrates the region where energy is permitted to enter or exit the budget, and the region where energy is merely redistributed. Figure \ref{fig:principal_sketch}(d) illustrates prototypical conserved triad pairs, with each pair denoted by a filled and open circle. The pairs are: $(l,n)$ and $(n,l)$, $(n-l,n)$ and $(n,n-l)$, and $(-l,n-l)$ and $(l-n,l)$. Together they form a six-triad conserved set. Depending on the frequencies $f_l$ and $f_n$, neighboring triads in figure \ref{fig:principal_sketch}(d) may merge into the degenerate sets in figure \ref{fig:principal_sketch}(e). Specifically, for $f_l=f_n$, the six-triad set degenerates into a four-triad conserved set, $(n,n)$, $(0,n)$, $(n,0)$, and $(-n,0)$, which are marked by circles; for $f_l=f_n/2$, the six-triad set instead degenerates into a three-triad conserved set, $(n/2,n)$, $(n,n/2)$, and $(-n/2,n2)$, marked by squares. Under certain conditions, the triads $(n,n)$ and $(-n/2,n/2)$, marked by half-filled symbols, each acts as a conserved repeated pair. Finally, figure \ref{fig:principal_sketch}(f) illustrates an arrangement of triads with positive (red) and negative (blue) energy transfers that indicates forward energy transfer from low to high frequencies, or large to small scales, while figure \ref{fig:principal_sketch}(g) indicates backward transfer from high to low frequencies, or small to large scales. We will expand on these concepts in the following sections.

\subsection{Inter-Triad Conservation of Spectral Energy}\label{subsec:conservation}

While the nonlinear term in the momentum equations can transfer and redistribute kinetic energy, it can neither produce nor remove net energy. The conservation of scale-to-scale kinetic energy transfer outlined in the following has been known at least as far back as the early text of \citet{Batchelor1953Cambridge}. Here, we closely follow the derivation presented by \citet{Barthel2022thesis}. It is common practice to assume that boundary contributions to $\hat{\mathcal T}_{l\to n}$ vanish, typically due to steady, no-slip, or periodic boundary conditions. However, for most numerical or experimental flow data where fluid enters and exits through the domain boundaries, these conditions do not hold. Therefore, we include the boundary term in our derivation. The significance of boundary contributions will become clear in the applications discussed in \S\ref{sec:applications}.

We define the integral energy transfer
\begin{align}
    \hat{\mathcal T}_{l\to n}  \equiv \int_{\Omega}\hat{T}_{l\to n}\dd{\vb*x} = -\int_{\Omega} \hat{\vb*u}_n^{\mathrm{H}}(\hat{\vb*u}_{n-l}\vdot\grad)\hat{\vb*u}_l \dd{\vb*x}
\end{align}
to quantify the energy transferred by the triad $(n-l,l,n)$. Integration by parts leads to
\begin{align}
    \hat{\mathcal T}_{l\to n} = \underbrace{- \int_{\Omega}\div{[(\hat{\vb*u}_n^{\mathrm{H}}\hat{\vb*u}_l)\hat{\vb*u}_{n-l}]}\dd{\vb*x} }_{F_{l\to n}} + \underbrace{ \int_{\Omega}\hat{\vb*u}_l^\mathrm{T}\div{(\hat{\vb*u}_n^*\hat{\vb*u}_{n-l}^\mathrm{T})}\dd{\vb*x} }_{R_2}.
\end{align}
The term $F_{l\to n}$ can be expressed using the divergence theorem as 
\begin{align}
   F_{l\to n} = -\oint_{\partial \Omega} \left( \hat{\vb*{u}}_{n}^{\mathrm{H}}  \hat{\vb*{u}}_{l} \right) \hat{\vb*{u}}_{n-l} \cdot \vb*{n} \, \dd l ,
\end{align}
which can be interpreted as the real part of the total bispectral density flux into the control volume, $\Omega$.
The term $R_2$ can be expanded using the product rule as 
\begin{align}
    R_2 =  \underbrace{ \int_{\Omega}\hat{\vb*u}_l^\mathrm{T}\hat{\vb*u}_n^{*}\div{\hat{\vb*u}_{n-l}}\dd{\vb*x} }_{R_3}   + \underbrace{ \int_{\Omega}\hat{\vb*u}_l^\mathrm{T}(\hat{\vb*u}_{n-l}\vdot\grad)\hat{\vb*u}_n^{*}\dd{\vb*x} }_{R_4}.
\end{align}
The term $R_3$ is zero due to incompressibility, $\div{\hat{\vb*u}}=0$. For real data, for which $\hat{\vb*{u}}_{l-n}=\hat{\vb*{u}}_{n-l}^*$ by the conjugate symmetry of the Fourier transform, and using the equation that
\begin{align}
    \hat{\mathcal T}_{n\to l} = - \int_{\Omega}\hat{\vb*u}_l^{\mathrm{H}}(\hat{\vb*u}_{l-n}\vdot\grad)\hat{\vb*u}_n\dd{\vb*x}  = - \int_{\Omega}\hat{\vb*u}_l^{\mathrm{H}}(\hat{\vb*u}_{n-l}^{*}\vdot\grad)\hat{\vb*u}_n\dd{\vb*x}  = -R_4^*,
\end{align}
we can obtain the pairwise relationships
\begin{subequations}\label{pairwise}
\begin{align}
    \hat{\mathcal T}_{l\to n} + \hat{\mathcal T}_{n\to l}^* &= F_{l \to n} = F_{n \to l} \\ \qand
    \hat{\mathcal T}_{l\to n}^\mathcal{R} + \hat{\mathcal T}_{n\to l}^\mathcal{R} &= F_{l \to n}^\mathcal{R} = F_{n \to l}^\mathcal{R},
\end{align}
\end{subequations}
where we defined the shorthand $(\cdot)^\mathcal{R}\equiv \mathcal{R}\{\cdot\}$. If the flux $F_{l \to n} = F_{n \to l}$ vanishes over the boundary $\partial \Omega$, the net real energy transfer is pairwise-conserved, $\hat{\mathcal T}_{l\to n}^\mathcal{R} + \hat{\mathcal T}_{n\to l}^\mathcal{R} = 0$.
Three such conserved pairs then form the six-triad conserved set,
\begin{align}\label{pairwise_three}
    (\hat{\mathcal T}_{l\to n}^\mathcal{R} + \hat{\mathcal T}_{n \to l}^\mathcal{R}) + (\hat{\mathcal T}_{n-l \to n}^\mathcal{R} + \hat{\mathcal T}_{n \to n-l}^\mathcal{R}) + (\hat{\mathcal T}_{-l \to n-l}^\mathcal{R} + \hat{\mathcal T}_{l-n\to l}^\mathcal{R}) = 0,
\end{align}
involving all permutations of the triplet $(n-l,l,n)$.


From the pairwise conservation expressed in equation (\ref{pairwise}), we can deduce that two types of triads form degenerate pairs and will have no integral energy transfer if the net flux is zero. The first type satisfies the component-wise equality $(n-l, l, n) = (l-n, n, l)$, which is equivalent to $l=n$ and corresponds to the magenta diagonal line in figure \ref{fig:principal_sketch}(a), which indicates spectral advection. Triads on either side of $l=n$ conserve energy in a pairwise fashion. The second type satisfies $(n-l, l, n) = -(l-n, n, l)$, equivalent to $l+n=0$, corresponding to the gray line, which is also the $f_k$-axis. Triad pairs on either side of $l+n=0$ are similarly conserved. In practice, the degenerate triads along the spectral advection line, $l=n$, may exhibit finite integral energy transfer, as the mean component of open flows tends to violate the zero-flux condition.

It follows from the pairwise and six-triad conservation properties in equations~\eqref{pairwise} and \eqref{pairwise_three} that the sum of all nonlinear transfer terms is zero, i.e.,
\begin{equation}\label{eq:global_conserve}
   \sum_{n=-\infty}^{\infty} \sum_{\substack{l\ne0, n}}\hat{\mathcal T}_{l\to n}^\mathcal{R} = 0,
\end{equation}
provided that the fluxes vanish over the domain boundary as before. Nonlinear energy transfer is thus both locally conserved by a limited set of triads and globally conserved by all triads. A visual interpretation is that the whole bispectral plane is a union of all nested hexagons. This reflects the well-known conservative property of the convective term. In summary, the integral energy transfer satisfies a hierarchy of conservative properties of increasing granularity: from global, to six-triad, to pairwise conservation. \Red{Energy is permitted to enter or leave the system only via mean production and linear advection, i.e., the green and magenta lines, respectively, in figure \ref{fig:principal_sketch}(a) (as well as through dissipation and other neglected terms), and is then distributed or scattered over the remainder of the bispectrum plane.}


\subsection{Modal Energy Budget and Directivity}\label{subsec:directivity}

Provided the flux, $F_{l\to n}$, is negligible, equation~\eqref{pairwise} is an expression of the conservative nature of spectral energy transfer \citep{kraichnan1959jfm}. The transfer of energy achieved by the triad $(n-l,l,n)$ is precisely balanced by the triad $(l-n,n,l)$. This observation motivates the recipient-donor interpretation, see \S\ref{sec:recdon}, of the energy transfer term, $\hat{\mathcal T}_{l\to n}$, in which the frequency components $f_l$ and $f_n$ are respectively identified as the donor and recipient of energy, and $f_{n-l}$ as the catalyst that neither donates nor receives energy, but merely mediates the transfer. This interpretation reframes the pairwise conservation in equation~\eqref{pairwise} as the statement that the energy transferred from $f_l$ to $f_n$ must be equal and opposite to the energy transferred from $f_n$ and $f_l$: a physically intuitive result. For any arbitrary donor-recipient pair, $f_l$ and $f_n$, the direction and quantity of the spectral transfer are given by the sign and magnitude, respectively, of $\hat{\mathcal T}_{l \to n}$ or $\hat{\mathcal T}_{n \to l}$. This allows the inter-scale propagation of energy across all frequencies to be systematically deduced from the convective modes and recipient modes of the decomposition. The notation $l\to n$ reflects this understanding. The donor and recipient frequencies are labeled in figure \ref{fig:principal_sketch}(a) in blue and red, respectively. \Red{Positive energy transfers in the region $f_n>|f_l|$ or, equivalently, negative energy transfers in the region $f_n<|f_l|$ signal the transport of energy from low to high frequencies, that is, a forward cascade. Conversely, an inverse cascade of energy towards lower frequencies can be readily identified by negative energy transfers in the region $f_n>|f_l|$ or, equivalently, positive energy transfers in the region $f_n<|f_l|$.}



Using the bilinearity of the inner product, we can expand the expected total energy transfer, abbreviated as $\hat{\mathcal T}_{l\to n}^\mathrm{avg} \equiv \mathrm{E}\qty{\hat{\mathcal T}_{l\to{n}}}$ in the following, in terms of the convective modes and recipient modes,
\begin{align}
\notag \hat{\mathcal T}_{l\to n}^\mathrm{avg} &= \mathrm{E}\qty{\hat{\mathcal T}_{l\to{n}}} = \mathrm{E}\qty{\ev*{\vb*{\hat c}_{l\to n},\vb*{\hat u}_n}} \\ \notag
&= \notag \mathrm{E}\ev{\sum_{j=1}^\infty b_j \hat{\vb*\psi}_{l\to n,j}, \sum_{j=1}^\infty a_j \hat{\vb*\phi}_{n,j}} \\ \notag
&= \mathrm{E}\underbrace{\ev*{b_1 \hat{\vb*\psi}_{l\to n,1}, a_1 \hat{\vb*\phi}_{n,1}}}_{\hat{\mathcal T}_{l\to n,1\to1}} + \mathrm{E}\underbrace{\ev*{b_2 \hat{\vb*\psi}_{l\to n,2}, a_1 \hat{\vb*\phi}_{n,1}}}_{\hat{\mathcal T}_{l\to n,2\to1}} + \mathrm{E}\underbrace{\ev*{b_1 \hat{\vb*\psi}_{l\to n,1}, a_2 \hat{\vb*\phi}_{n,2}}}_{\hat{\mathcal T}_{l\to n,1\to2}} + \mathrm{E}\underbrace{\ev*{b_2 \hat{\vb*\psi}_{l\to n,2}, a_2 \hat{\vb*\phi}_{n,2}}}_{\hat{\mathcal T}_{l\to n,2\to2}} + \ldots \\ \notag
&= \mathrm{E}\qty{b_1a_1^*}\ev*{\hat{\vb*\psi}_{l\to n,1}, \hat{\vb*\phi}_{n,1}} + \mathrm{E}\qty{b_2a_1^*}\ev*{\hat{\vb*\psi}_{l\to n,2}, \hat{\vb*\phi}_{n,1}} + \mathrm{E}\qty{b_1a_2^*}\ev*{\hat{\vb*\psi}_{l\to n,1}, \hat{\vb*\phi}_{n,2}} + \mathrm{E}\qty{b_2a_2^*}\ev*{\hat{\vb*\psi}_{l\to n,2}, \hat{\vb*\phi}_{n,2}} + \ldots \\ \notag
&= \sigma_1\ev*{\hat{\vb*\psi}_{l\to n,1}, \hat{\vb*\phi}_{n,1}} + \sigma_2\ev*{\hat{\vb*\psi}_{l\to n,2}, \hat{\vb*\phi}_{n,2}} + \ldots \\ 
&= \sum_{j=1}^\infty \sigma_j \ev*{\hat{\vb*\psi}_{l\to n,j}, \hat{\vb*\phi}_{n,j}} = \sum_{j=1}^\infty {\hat{\mathcal T}^\mathrm{avg}_{l\to n,j}} \label{eqn:Tavg},
\end{align}
where we have made use of the property that the expansion coefficients are uncorrelated, see equation \eqref{eq:coeffsUncorrelated}. Each term in the expansion, ${\hat{\mathcal T}^\mathrm{avg}_{l\to n,j}} = \sigma_j \ev*{\hat{\vb*\psi}_{l\to n,j}, \hat{\vb*\phi}_{n,j}}$, represents the average integral modal energy transfer from the $j$th convective mode to the $j$th recipient mode at frequency $f_n$. It follows from the uncorrelatedness of the expansion coefficients that, on average, no energy is exchanged between the $i$th and $j$th modes for $i\neq j$. For brevity, we therefore only use the subscript $j$ to represent these transfers.


In general, the conservation of energy described in \S \ref{subsec:conservation} is satisfied in aggregate for $\hat{\mathcal T}_{l\to{n}}^\mathrm{avg}$, not for an individual modal transfer, $\hat{\mathcal T}_{l\to n,j}^\mathrm{avg}$. However, $\hat{\mathcal T}_{l\to{n}}^\mathrm{avg}$ will be well-represented by $\hat{\mathcal T}_{l\to n,1}^\mathrm{avg}$ if the singular values, $\sigma_j$, decay rapidly, that is, if the leading mode pair captures the majority of the total convective-recipient covariance. In such a case, $\hat{\mathcal T}_{l\to n,1}^\mathrm{avg}$ should approximately be conserved, as we will demonstrate in \S \ref{sec:cylinder}. For the remainder of this work, we will thus focus on the transfer between the leading convective mode and recipient mode.


We can further relate the magnitude of the expected integral modal transfer, ${\hat{\mathcal T}^{\mathrm{avg}}_{l\to n,j}}$, and that of its real part, $\mathcal{R}\left\{\hat{\mathcal T}^{\mathrm{avg}}_{l\to n,j}\right\}\equiv \hat{\mathcal T}^{\mathrm{avg},\mathcal{R}}_{l\to n,j}$, to the mode bispectrum, $\sigma_j$, by writing
\begin{align}\label{eq:energyTransferBound}
  \qty|\hat{\mathcal T}^{\mathrm{avg},\mathcal{R}}_{l\to n,j}| = \qty|\mathcal{R}\left\{\sigma_j\ev*{\hat{\vb*\psi}_{l\to n,j}, \hat{\vb*\phi}_{n,j}}\right\}| = \sigma_j \underbrace{\qty|\mathcal{R}\Bqty{\ev*{\hat{\vb*\psi}_{l\to n,j}, \hat{\vb*\phi}_{n,j}}}|}_{\in \qty[0,1] },
\end{align}
where we have invoked the Cauchy-Schwarz inequality in the last step. In other words, the integral modal transfer is proportional to and bounded by the singular value, with the scaling factor given by the degree of spatial alignment between the convective and recipient modes.

From expanding the expected integral modal transfer 
in terms of the recipient and convective modes according to equation (\ref{eqn:Tavg}),
\begin{equation}
\hat{\mathcal T}_{l\to n,j}^\mathrm{avg} =
 \int_f \sigma_j \hat{\vb*\phi}_{n,j}^\mathrm{H} \hat{\vb*\psi}_{l\to n,j} \dd{\vb*x} = \int_f \mathrm{E}\qty{\hat{T}_{l\to n,j}} \dd{\vb*x},
\end{equation}
it becomes evident that its value is determined by integrating the mode product $\hat{\vb*\phi}_{n,j}^\mathrm{H} \hat{\vb*\psi}_{l\to n,j}$, weighted by $\sigma_j$, over the domain of interest, $f$. This motivates the definition of the integrand of $\hat{\mathcal T}_{l\to n,j}^\mathrm{avg}$, 
\begin{equation} \label{eqn:transferfield}
\hat{\tau}_{l\to n,j} \equiv  \sigma_j\hat{\vb*\phi}_{n,j}^\mathrm{H} \hat{\vb*\psi}_{l\to n,j} ,   
\end{equation}
as the modal \emph{transfer field}. We note that $\hat{\tau}_{l\to n,j}$ is a complex quantity.

\Red{Owing to the conservation properties described in \S\ref{subsec:conservation}, the expected integral modal transfer, $\hat{\mathcal T}^{\mathrm{avg}}_{l\to n,j}$, has a principal region that spans the sector $f_n \ge |f_l|$, between the gray and magenta lines in figure \ref{fig:principal_sketch}(a). All other integral transfers can be recovered from the principal region via pairwise conservation. On the other hand, the modes, $\hat{\vb*\psi}_{l\to n,j}$ and $\hat{\vb*\phi}_{n,j}$, and transfer field, $\hat{\tau}_{l\to n,j}$, do not obey conservation, as will become clear in the examples (\S\ref{sec:applications}). Their principal region is thus the entire top-half plane in figure \ref{fig:principal_sketch}(a).}

\section{Applications}\label{sec:applications}

\begin{figure}
    \centering
    \includegraphics[width=0.6\linewidth]{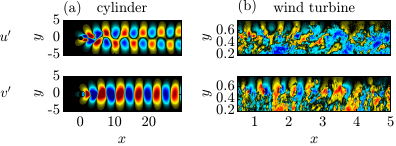}
    \caption{Instantaneous fluctuating streamwise (first row) and transverse (second row) velocities of the two examples considered in \S\ref{sec:applications}: (a) cylinder wake; (b) wind turbine wake. Normalized fields (\,\fcolorbox{blue}{blue}{\rule{0pt}{3pt}\rule{3pt}{0pt}}\fcolorbox{black}{black}{\rule{0pt}{3pt}\rule{3pt}{0pt}}\fcolorbox{red}{red}{\rule{0pt}{3pt}\rule{3pt}{0pt}} blue, black, red, $q/|\max{\{q\}}|\in[-0.5,0.5 ]$) are shown. In (b), for clarity the $x$- and $y$-axes are not drawn to scale. In what follows, all contours use the same axes limits and colorscales as (a) for the cylinder wake or (b) for the turbine wake.}
    \label{fig:overview}
\end{figure}

\begin{table}[!htb]
      \centering
    \renewcommand{\arraystretch}{1.0}
        \begin{tabular}{|c|c|c|c|c|c|c|c|c|c|c| }
      \cline{1-11}
    Case & \S & $\mathrm{Re}$  & $f_0$  & ${\Delta t U_\infty}/{D}$  & $N_x$ & $N_y$ & $N_t$ & $N_f$ & $N_\mathrm{ovlp}$ & $N_\mathrm{blk}$ \\
 \hline 
 \hline 
 Cylinder wake  DNS \citep{chu2023rbf}&    \S \ref{sec:cylinder}  & $100$ & $0.167$   & 0.34  & 230 & 101 & 600 & 150 & 75  & 7  \\
 \hline
  Wind turbine TR-PIV \citep{BiswasBuxton2024jfm,BiswasBuxton2024energy} &  \S \ref{sec:windturbine} & $\approx 4\times 10^4$  & $1.87$  & 0.01 & 769 & 196 & 5456 & 321 & 161 & 33 \\ 
 \hline
        \end{tabular}
        \caption{Overview of datasets and spectral estimation parameters.}
    \label{flow_summary}
\end{table}

We demonstrate the decomposition and modal energy flow analysis using two different flow cases. The first case involves numerical simulation data of the wake behind a cylinder at $\mathrm{Re}=100$, a canonical example of unsteady laminar flow with well-established nonlinear dynamics. The second case uses PIV data from a wind turbine wake, illustrating the application of TOD to turbulent flows, even in the presence of measurement noise. Instantaneous snapshots of both flows are shown in figure \ref{fig:overview}, while the data sizes, flow parameters, and spectral estimation settings for both cases are summarized in table \ref{flow_summary}.

In each case, we define a fundamental frequency, $f_0$, which is used to normalize the frequency axes. This normalization simplifies the identification of harmonic frequencies, allowing us to refer to them by their integer multiples of $f_0$, thus making the bispectral plots easier to interpret. For the cylinder wake, the vortex-shedding frequency is used as the fundamental frequency, and for the wind turbine wake, we use the rotor rotational frequency. For the first example, we systematically present the method's primary outcomes and illustrate the interpretation of the results. The second example provides a comprehensive analysis, focusing on the physical explanation of a complex technical flow.
 

\subsection{Cylinder Wake }\label{sec:cylinder}



We start with the cylinder wake, a classic example of a canonical unsteady flow characterized by well-understood nonlinear dynamics and commonly used in reduced-order modeling \citep{deane1991low,ma2002low,NOACK_2003,jin2021energy}. Specifically, we analyze the flow around a cylinder at a Reynolds number of $\mathrm{Re} = 100$, based on the cylinder diameter and free-stream velocity.
The direct numerical simulation (DNS) data for this case is obtained using the PHS+poly RBF-FD implementation of the fractional-step, staggered-grid incompressible Navier-Stokes solver developed by \citet{chu2023rbf}. The cylinder is positioned at the origin with a diameter of 1. The time step in the simulation is adjusted to ensure that a Fourier block of size $N = 150$ captures multiple complete flow cycles. The fundamental vortex shedding frequency is $St_{0} = 0.1673$ \citep{williamson1988defining, barkley2006linear, jiang2017strouhal}. The flow fields are interpolated onto a Cartesian mesh within the computational domain defined as $x, y \in [-5, 29.35] \times [-5, 5]$.

The first outcomes of TOD are two bispectral plots: one depicting the mode bispectrum and the other illustrating the modal energy budget. These plots provide a comprehensive overview of the triadic phase coupling and energy transfer across frequencies, respectively, and are presented in \S\ref{sec:cylinder_bispectra}. The second outcome consists of three physically interpretable fields: the recipient modes, convective modes, and transfer fields. Examples of these fields and their interpretation are detailed in \S\ref{sec:cylinder_bispectra}.

The mode bispectrum, along with the recipient and convective mode pairs, is computed via the SVD of the two-point cross-bispectral covariance matrix in equation (\ref{Sfq_SVD}). The modal energy budget, representing the expected total energy transfer in the bispectral plane, and the corresponding transfer fields are computed from the modes according to definitions (\ref{eqn:Tavg}) and (\ref{eqn:transferfield}), respectively. In both examples, we show that focusing on the leading, or optimal, donor-recipient pair is sufficient.

\begin{figure}[ht]
    \centering
   \begin{tikzpicture}
    \pgftext{  \includegraphics[trim = 0mm 0mm 0mm 0mm, clip, width=0.6\textwidth]{ 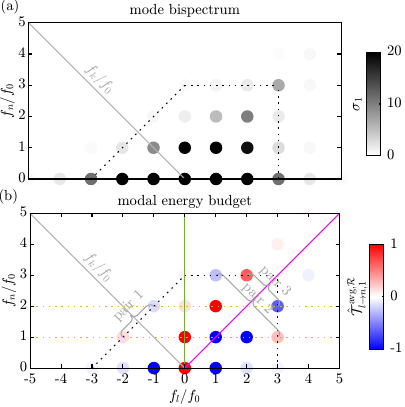}}
     \node[red] at (0.05,-2.4) {\ref{fig:Cylinder-tod_mod_complex}};
      \node[orange] at (-3.5,-2.4){ \ref{fig:Cylinder-tod_mod-same-recipient}(b)};
     \node[orange] at (-3.5,-3.1){ \ref{fig:Cylinder-tod_mod-same-recipient}(a)};
     \node[gray, rotate=45] at (-1,-1.8){ \ref{fig:Cylinder-tod_mod}(a,b)};
          \node[gray, rotate=-45] at (2.3,-3.2){ \ref{fig:Cylinder-tod_mod}(c,d)};        \node[gray, rotate=-45] at (2.55,-2.75){ \ref{fig:Cylinder-tod_mod}(e,f)};
    \end{tikzpicture}
    \caption{
    Scatter plots of the leading mode bispectrum (a) and modal energy budget (b)
    of the cylinder wake at Re=100. Colored lines mark linear advection (magenta), mean production (green), and the
    catalyst frequency axis (gray); see figure \ref{fig:principal_sketch}(a). Energy-conserving pairs involving the fundamental vortex shedding frequency, $f_0$, are also marked. Numbers and letters indicate figures depicting the corresponding modes.
    }
    \label{fig:Cylinder-tod_spec}
\end{figure}
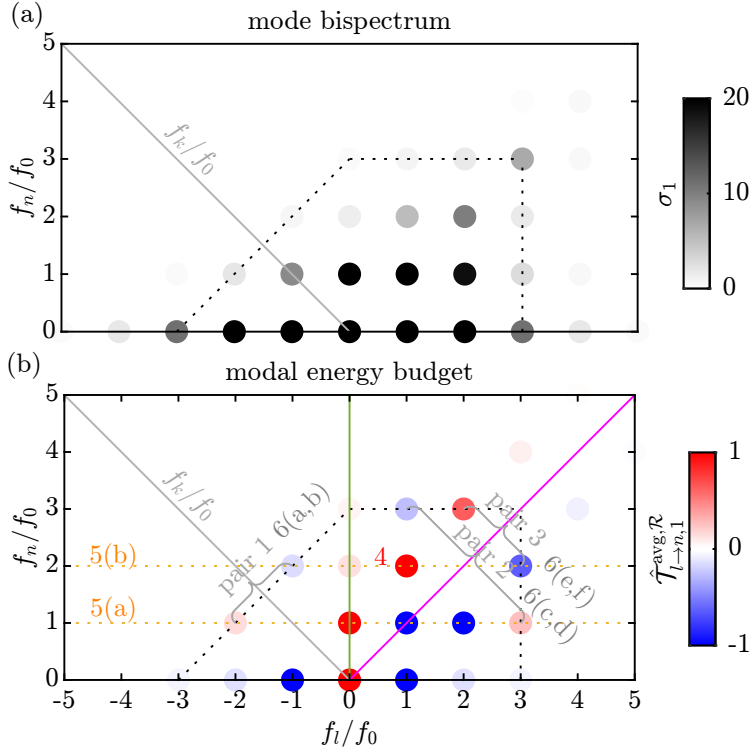

\subsubsection{Mode bispectrum and modal energy budget} \label{sec:cylinder_bispectra}

To identify the most active triadic interactions, we present the mode bispectrum, which shows the optimal convective-recipient covariance $\sigma_1$, and the corresponding (real part of the) modal energy transfer $\hat{\mathcal T}_{l\to n,1}$, in panels \ref{fig:Cylinder-tod_spec}(a) and \ref{fig:Cylinder-tod_spec}(b), respectively. In both panels, local maxima occur at integer multiples of the fundamental vortex shedding frequency, underscoring the importance of these harmonics. Triads with negligible amplitudes are omitted for clarity. Each panel emphasizes different aspects of the triadic interactions and should be analyzed together for a more comprehensive understanding. The mode bispectrum effectively identifies the triads with the strongest phase coupling but does not provide a quantitative measure of the total energy transfer associated with these triads. Conversely, the modal energy budget quantifies these energy transfers but does not directly indicate the strength of the phase couplings. Interestingly, some triads prominent in the bispectrum, such as \( (f_l, f_n)/f_0 = (-1,1) \) and \( (2,2) \), show negligible integral modal energy transfer, suggesting that strong phase coupling can occur alongside zero net energy exchange between the convective terms and the recipient modes in these triads.


The most significant triads are observed along the lines \( f_n = 0 \) and \( f_n = f_0 \) in figure \ref{fig:Cylinder-tod_spec}~(a). The former indicates that the spatial deformation of the mean flow field is primarily driven by the self-interactions between the harmonics and their corresponding conjugates. Meanwhile, the triads along the line \( f_n = f_0 \) emphasize the triadic activity at the fundamental vortex-shedding frequency. Additionally, the advection of harmonics is evident along the line \( l = n \). Other significant triads, such as the sum interaction of the fundamental instability with itself, contributing to the second harmonic (i.e., \( (f_l, f_n)/f_0 = (1, 2) \)), are also well-captured. 

The residual mode bispectrum, which contains all higher modes (not shown), lacks such structure and is negligible compared to the optimal convective-recipient covariance $\sigma_1$, with a covariance fraction of \( \sum_{j = 2}^7 \left(\sum_n \sum_l \sigma_j\right) / \sum_{j = 1}^7 \left(\sum_n \sum_l \sigma_j\right) \approx 10^{-8} \). This stark separation of singular values indicates that it is sufficient to focus solely on the leading singular value.

Figure \ref{fig:Cylinder-tod_spec}~(b) provides a detailed analysis of the modal energy budget and its directional characteristics for each triad. As expected, the largest energy transport is observed for the self-advection at the origin, \( (f_l, f_n)/f_0 = (0,0) \). Additionally, several active triads with significant energy transfer are found along the line \( f_n = f_0 \), where the recipient frequency matches the fundamental frequency (lower orange dotted line). It is evident that this frequency gains energy from the mean flow through production and subsequently contributes to the superharmonic via a forward energy cascade. At the nodes \( (f_l, f_n)/f_0 = (-2,1) \) and \( (3,1) \), we observe an inverse energy cascade, where energy transfers from higher to lower frequencies. Interestingly, a negative energy budget is present at \( (f_l, f_n)/f_0 = (1, 1) \), indicating a deceleration of vortex shedding when advected by the mean flow. This phenomenon is further illustrated in figure \ref{fig:Cylinder-tod_mod-same-recipient}(d), showing two symmetrically positioned deceleration lobes downstream of the cylinder. When the recipient frequency is the superharmonic (i.e., \( f_n/f_0 = 2 \)), it mainly gains energy from the fundamental frequency and contributes to the third harmonic, demonstrating a successive transfer of energy from lower to higher frequencies. A clear forward energy cascade is apparent in the sequence of positive energy transfers above the \( f_n = f_l \) line, as well as the negative energy transfers below it with matching magnitudes. Furthermore, the modal energy budget is anti-symmetric about the \( f_n = f_l \) line in the right quadrant, implying that the amount of energy transferred from the donor to the recipient equals the amount received by the recipient. This confirms that there is no net energy gain or loss during the transfer process. In the left quadrant, the anti-symmetry follows the line \( f_l + f_n = 0 \). These observations exemplify the property of pairwise conservation described in \S \ref{subsec:conservation} and illustrated in figure~\ref{fig:principal_sketch}. 
As an example, we highlight three easily identifiable conservation pairs located on the same hexagon. Setting $l=2$ and $n=3$ in equation (\ref{pairwise_three}) 
shows that these three conservation pairs constitute the six-triad conserved set
\begin{align}\label{pairwise_examples}
    \underbrace{(\hat{\mathcal T}_{2\to 3}^\mathcal{R} + \hat{\mathcal T}_{3 \to 2}^\mathcal{R})}_{\text{pair 3}} + \underbrace{(\hat{\mathcal T}_{1 \to 3}^\mathcal{R} + \hat{\mathcal T}_{3 \to 1}^\mathcal{R})}_{\text{pair 2}} + \underbrace{(\hat{\mathcal T}_{-1 \to 2}^\mathcal{R} + \hat{\mathcal T}_{-2\to 1}^\mathcal{R})}_{\text{pair 1}} = 0,
\end{align}
highlighting how both individual pairs and the six-triad set maintain a zero net energy balance. For $l=0$ and $n=3$, we obtain the degenerate set 
\begin{align}\label{complete_degenerates}
    \underbrace{(\hat{\mathcal T}_{0\to 3}^\mathcal{R} + \hat{\mathcal T}_{3 \to 0}^\mathcal{R})}_{\text{pair 3}} + \underbrace{(\hat{\mathcal T}_{3 \to 3}^\mathcal{R} + \hat{\mathcal T}_{3 \to 3}^\mathcal{R})}_{\substack{\text{degenerate pair}\\ \text{(mean advection)}}} + \underbrace{(\hat{\mathcal T}_{0 \to 3}^\mathcal{R} + \hat{\mathcal T}_{-3\to 0}^\mathcal{R})}_{\text{pair 1}} = 0,
\end{align}
which contains two repeated triads, $\hat{\mathcal T}_{3 \to 3}^\mathcal{R}$ and $\hat{\mathcal T}_{0\to 3}^\mathcal{R}$, and therefore only four unique triads. Unlike the triad associated with the advection of the fundamental, $\hat{\mathcal T}_{1\to 1}^\mathcal{R}$, the triad associated with the advection of the third harmonic, $\hat{\mathcal T}_{3 \to 3}^\mathcal{R}$, has near-zero net energy transfer. This leaves nine triads involving the mean and the first three harmonics with non-zero net energy transfer out of the two sets described by equations (\ref{pairwise_examples}) and (\ref{complete_degenerates}). When taken together, these nine triads complete the conservation of energy on the hexagon indicated by the dotted line.


\subsubsection{Recipient modes, convective modes, and transfer fields} \label{sec:cylinder_modes}

\begin{figure}
    \centering
    \includegraphics[trim = 0mm 0mm 0mm 0mm, clip, width=1\textwidth]{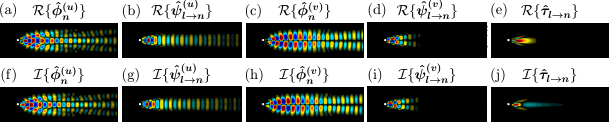}
    \caption{
    The leading recipient modes, convective modes, and transfer fields for the donor-recipient pair $(f_l,f_n)/f_0=(1,2)$: (a-e) real parts; (f-j) imaginary parts; (a,b,f,g) streamwise components; (c,d,h,i) transversal components.
   }
    \label{fig:Cylinder-tod_mod_complex}
\end{figure}

In addition to the mode bispectrum and the modal energy budget—both evaluated in an integral sense—TOD analysis provides physically interpretable fields, offering deeper insights into the spatial characteristics of triadic interactions. As a representative example, we visualize the recipient modes, convective modes, and transfer fields for \( (f_l, f_n)/f_0 = (1, 2) \) in figure \ref{fig:Cylinder-tod_mod_complex}. The real and imaginary components of the recipient and convective modes display identical flow patterns with a constant streamwise phase shift of approximately \(\pi/2\) radians, as expected for nondispersive structures that simply convect downstream. As noted in \S \ref{sec:recdon}, both fields evolve coherently in time at the recipient frequency. The real component of the transfer field shows that an oval region downstream of the cylinder predominantly gains energy. Therefore, in the following analysis, we will focus on the real components of these fields and the streamwise components of the recipient and convective modes.

\begin{figure}[ht]
    \centering
    \includegraphics[trim = 0mm 0mm 0mm 0mm, clip, width=1\textwidth]{ 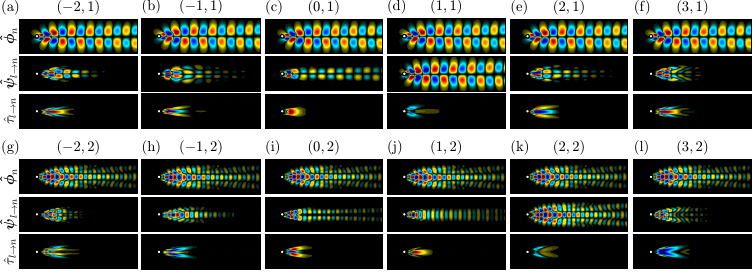}
    \caption{
    The real part of the leading stream-wise recipient and convective modes, and transfer fields of the cylinder wake for the same recipient frequencies: (a-f) $f_n/f_0=1$, harmonic; (g-l) $f_n/f_0=2$, superharmonic. 
    }
    \label{fig:Cylinder-tod_mod-same-recipient}
\end{figure}

Figure \ref{fig:Cylinder-tod_mod-same-recipient} presents the leading modes for two recipient frequencies: the harmonic and superharmonic. The leading recipient modes remain consistent across different donor frequencies, while the convective modes vary for the same recipient frequency. Together, these observations emphasize the collective manner in which triads interact to produce the observed flow structures. It is noteworthy that the transfer fields exhibit compact support, even though the convective and recipient modes are active far downstream of the cylinder, extending beyond the computational domain. This suggests that the most critical triadic interactions occur close to the cylinder. The compact support of the interaction regions aligns with findings from experimental structural perturbation analysis \citep{strykowski1990formation} and prediction-based wavemaker analysis \citep{giannetti2007structural,marquet2008sensitivity}. The latter, a linear approach, identifies regions in the flow with the strongest localized feedback where dominant instability mechanisms are active. Going beyond linear analysis, the transfer fields now identify these regions as the nonlinear most active.


\begin{figure}
    \centering
    \includegraphics[trim = 0mm 0mm 0mm 0mm, clip, width=1\textwidth]{ 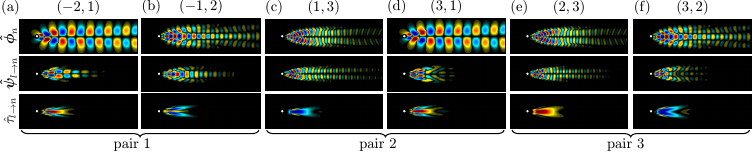}
    \caption{
    Same as figure \ref{fig:Cylinder-tod_mod-same-recipient} but for the 3 conserved pairs shown in figure \ref{fig:Cylinder-tod_spec}(b) and equation (\ref{pairwise_examples}).}
    \label{fig:Cylinder-tod_mod}
\end{figure}

Figure \ref{fig:Cylinder-tod_mod} illustrates the leading fields for the three conservative pairs highlighted in panel \ref{fig:Cylinder-tod_spec}(b) and equation (\ref{pairwise_examples}).
Each pair of transfer fields reveals two distinct structures, but their integral sum remains zero, demonstrating the balanced contributions from both components. This confirms that energy is spatially redistributed through nonlinear interactions within each pair, while the total energy remains conserved. Thus, analyzing the entire transfer field structures, rather than just their integral sums, is crucial for fully understanding the physical mechanisms underlying these nonlinear interactions. Pairs 1 and 2 display the flow fields associated with the inverse energy cascade. Interestingly, the energy transfer fields show a similar pattern when the recipient frequencies are lower than the donor frequency but differ significantly when the recipient frequency is higher. This suggests that similar mechanisms govern the energy transfer from the negative superharmonic and the third harmonic back to the fundamental frequency.





\subsection{Wind Turbine Wake}\label{sec:windturbine}
In the second application, we demonstrate the utility of the new orthogonal decomposition for the identification of dominant triads in experimental data of turbulent flows. We choose the example of the turbulent wake of a model wind turbine from \citet{BiswasBuxton2024jfm,BiswasBuxton2024energy}, with a Reynolds number of 40000 based on the rotor diameter and freestream velocity. The turbine wake flow exemplifies real-world data of mixed broadband-tonal turbulent flows, whose complex dynamics exhibits both stochastic and deterministic behaviors, in the presence of noise and other measurement artifacts. The data were captured using time-resolved particle image velocimetry (TR-PIV), and correspond to experiment 1A in \citet{BiswasBuxton2024energy}. The field of view (FOV) of the PIV is aligned with the axis of the turbine tower, and spans $x/D\in[0.5,5]$ and $y/D\in[-0.35,0.75]$ in the streamwise and transverse directions, respectively, where $D$ is the diameter of the rotor. The rotor revolves at a tip speed ratio of $\lambda=\pi St_0\approx6$, where $St_0=f_0D/U_\infty=1.87$ is the rotational frequency and $U_\infty$ is the freestream velocity. The time series consists of 5456 snapshots separated by a constant spacing of $\Delta tU_\infty/D=0.01$, and spans 102 revolutions. The period of each revolution is given by $TU_\infty/D=1/St_0=0.535$. The three-bladed rotor also imposes a blade-passing frequency (BPF) of $3St_0=5.61$ on the flow. \citet{BiswasBuxton2024energy} previously analyzed this flow using the technique of optimal mode decomposition (OMD; \citep{WynnEtAl2013JFM}), itself an extension to dynamic mode decomposition (DMD; \citep{Schmid2010JFM}). They extracted coherent structures using OMD, then projected the kinetic energy budget equation onto the dominant structures, finding significant energy transfer among $St_0$, $3St_0$, and their harmonic peaks via triadic interactions. The turbine wake is thus ideal for assessing TOD, which, distinct from OMD and DMD, directly and optimally accounts for triads.

\begin{figure}
    \centering
    \begin{tikzpicture}
    \pgftext{\includegraphics[width=1\textwidth]{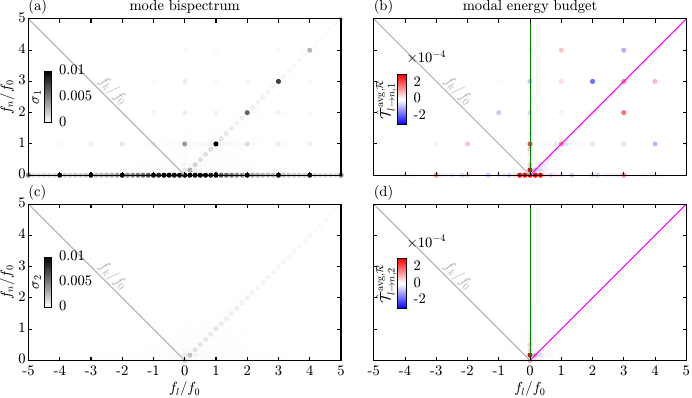}}
    \node at (4.05,1.4) {\ref{fig:turbine-tod_mode}(a)};
    \node at (4.05,2.2) {\ref{fig:turbine-tod_mode}(b)};
    \node at (4.05,3.0) {\ref{fig:turbine-tod_mode}(c)};
    \node at (5.6,3.0) {\ref{fig:turbine-tod_mode}(d)};
    \node at (6.35,2.2) {\ref{fig:turbine-tod_mode}(e)};
    \node at (6.35,3.0) {\ref{fig:turbine-tod_mode}(f)};
    \end{tikzpicture}
    \caption{Mode bispectra (a,c) and modal energy budgets (b,d) of the turbine wake for the leading (a,b) and first suboptimal (c,d) modes.}
    \label{fig:turbine-tod_spec}
\end{figure}
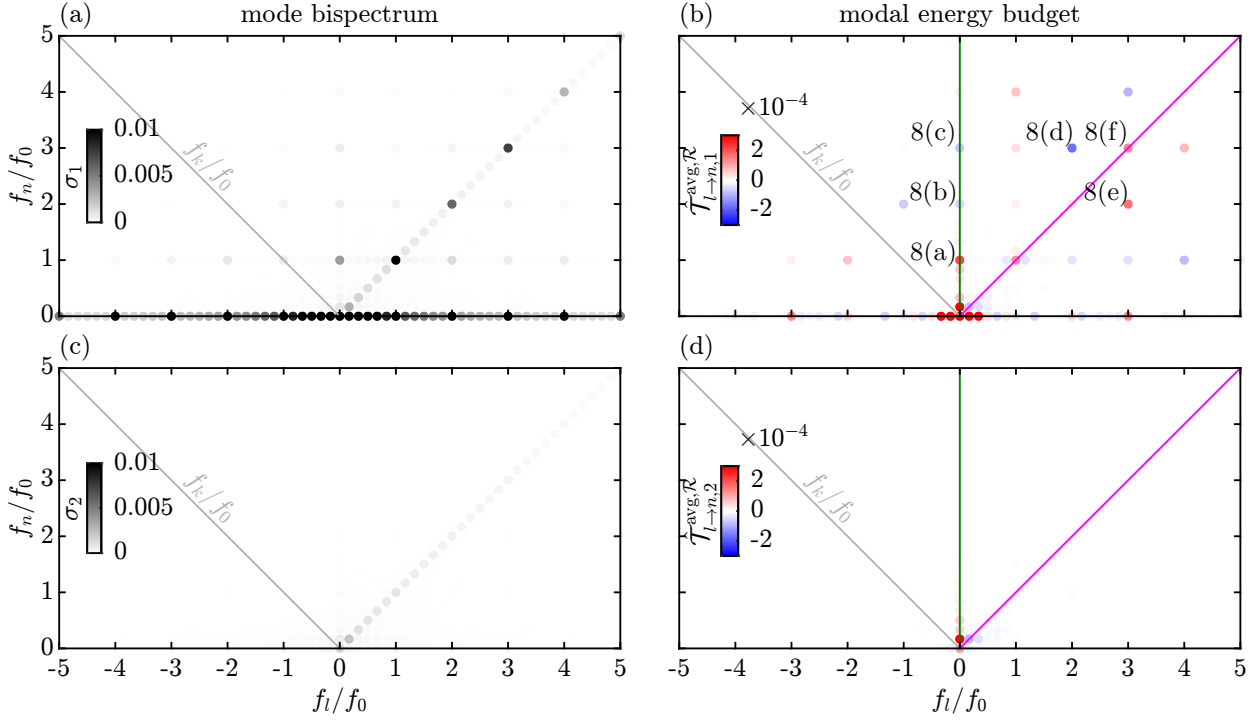
For spectral estimation, we choose a block size of $N_f=321$, equivalent to six rotational periods, and use a rectangular window. As with the cylinder wake in \S \ref{sec:cylinder}, this combination of parameters ensures that the discrete spectral energy distribution of the turbine wake is accurately captured. Figure~\ref{fig:turbine-tod_spec} shows the mode bispectra and modal energy budgets corresponding to the first and second TOD modes. The leading singular values in figure~\ref{fig:turbine-tod_spec}(a), summed over the bispectrum plane, account for a convective-recipient covariance fraction of $(\sum_n\sum_l \sigma_1)/(\sum_{j=1}^{N_b}\sum_n\sum_l \sigma_j) = 63\%$, thus confirming the large separation between the leading and suboptimal covariances in the turbulent turbine wake. By contrast, the first suboptimal singular values in figure~\ref{fig:turbine-tod_spec}(c) display substantially reduced magnitudes, except at non-harmonic frequencies along $f_n=f_l$. This is indicative of negligible nonlinear interactions in the suboptimal modes. The integral modal energy budgets in figure~\ref{fig:turbine-tod_spec}(b) and \ref{fig:turbine-tod_spec}(d) for the first and second modes, respectively, bolster this claim. Whereas the leading mode exhibits significant transfer of energy over the entirety of the bispectrum, including mean production ($f_l=0$), linear advection ($f_n=f_l$), and nonlinear transfer, the suboptimal mode shows only mean production at low frequencies, and no other sources of significant energy transport. TOD modes are thus an efficient representation of the triadic linear and nonlinear dynamics of the flow.

The leading mode bispectrum and integral modal energy budget reveal a grid-like pattern of local maxima, which identify triads made up of the rotational frequency, $f_0$, and its higher harmonics. Unlike the cylinder wake, in the turbine wake, the optimal convective-recipient covariance, $\sigma_1$, and the optimal real integral modal energy transfer, $\hat{\mathcal T}^{\mathrm{avg},\mathcal{R}}_{l\to n,1}$, both display finite values at non-harmonic frequencies, which reflects the turbulent nature of the flow. However, the magnitudes of both statistics at non-harmonic frequencies tend to be small, so we focus on the harmonics. Because the turbine wake exhibits large singular value separation, we expect---and confirm---that the leading modes of these dominant triads approximately recover the conservation principles described in \S\ref{subsec:conservation} and exemplified by the laminar cylinder wake in \S\ref{sec:cylinder}. With the exception of triads along the MKE line, $f_n=0$, mean production line, $f_l=0$, and linear advection line, $f_n=f_l$, the modal energy budget largely obeys pairwise conservation, thus facilitating the interpretation of energy flow from $f_l$ to $f_n$ previously introduced in \S\ref{subsec:directivity}.

In the leading mode bispectrum in figure~\ref{fig:turbine-tod_spec}(a), the most dominant triads are found along the $f_n=0$ axis. Each of these triads corresponds to the convective term $\vb*{\hat{c}}_{l\rightarrow 0}$ within the RANS equations \eqref{eq:RANS}. They thus conspire to deform the turbulent mean flow through the action of the Reynolds stress, $\overline{(\vb*u'\vdot\grad)\vb*u'}$.
The diagonal line representing $f_n=f_l$, or equivalently, $f_k=0$, also displays active triad interactions. Along this line, the local maxima correspond to the term $\vb*{\hat{c}}_{n\rightarrow n}$ in the spectral momentum equation~\eqref{spec_mom_eqn}, with $f_n$ equal to the harmonic frequencies. These triads arise due to advection of the turbine blade tip vortices by the mean flow. The dominance of these triads in the mode bispectrum reflects the convective nature of the turbine wake. The most active of these triads are $(f_l,f_n)/f_0=(1,1)$ and $(3,3)$. They couple the mean flow to the exogenous forcing, i.e., the rotational and blade-passing frequencies, $f_0$ and $3f_0$, respectively. 
Nonlinear triads, which do not involve the mean flow as one frequency component, are also clearly identified in the mode bispectrum, but are considerably weaker than those that do.

The leading integral modal energy budget in figure~\ref{fig:turbine-tod_spec}(b) is generally in good agreement with the energy budget and directivity reported by \citet{BiswasBuxton2024energy} based on OMD modes. By recognizing mean production and linear advection as special cases of triadic energy transfer, the modal budget enables us to directly compare the importance of these linear mechanisms to the nonlinear transfer between specific donor-recipient frequency pairs. The rotational frequency, $f_n/f_0=1$, receives most of its energy from $f_l=0$ via mean production, $(f_l,f_n)/f_0=(0,1)$. The second harmonic, $f_n/f_0=2$, receives most of its energy from the blade-passing frequency, $f_l/f_0=3$, via the nonlinear transfer $(f_l,f_n)/f_0=(3,2)$. The blade-passing frequency, in turn, acts mainly as a donor of energy back to the second harmonic via $(f_l,f_n)/f_0=(2,3)$. In \citet{BiswasBuxton2024energy}, it was shown that linear advection is the main contribution to the energy budget of $f_n/f_0=3$, as the \emph{net} nonlinear transfer to this frequency is comparatively small. Here, we see that the \emph{individual} nonlinear transfer, $(f_l,f_n)/f_0=(2,3)$, in fact dominates. In other words, on a granular level, the overall spectral energy budget at $f_n/f_0=3$ is strongly dependent on nonlinear interactions. 

It should come as no surprise that the modal energy budget of the cylinder wake differs significantly from that of the turbine wake. Whereas the former is characterized by the natural instability of vortex shedding, the latter is forced exogenously by the rotating blades. In the turbine wake, at every harmonic frequency, $f_n/f_0=1,2,\ldots$, linear advection makes a positive contribution to the energy budget. This observation stands in contrast to the budget of the cylinder wake in figure \ref{fig:Cylinder-tod_spec}, where linear advection removes energy from $f_n/f_0=1$, and preserves the energy of the higher harmonics. Furthermore, in the cylinder wake, mean production injects energy into all harmonics. In the turbine wake, on the other hand, the mean flow contributes energy to $f_n/f_0=1$ while extracting energy from $f_n/f_0=2, 3$. More generally, nonlinear interactions play a greater role in the overall dynamics of the cylinder wake than they do in the turbine wake.

\begin{figure}
    \centering
    \includegraphics[width=.8\textwidth]{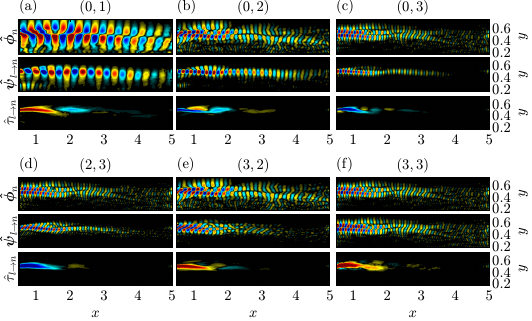}
    \caption{The real part of the leading streamwise recipient modes (first and fourth rows), streamwise convective modes (second and fifth rows), and transfer fields (third and sixth rows) of the turbine wake for different donor-recipient pairs, $(f_l/f_0,f_n/f_0)$: (a) $(0,1)$; (b) $(0,2)$; (c) $(0,2)$; (d) $(2,3)$; (e) $(3,2)$; (f) $(3,3)$. 
    }
    \label{fig:turbine-tod_mode}
\end{figure}
Figure \ref{fig:turbine-tod_mode}(a--c) shows the leading recipient modes, $\hat{\vb*\phi}_{n,1}$, convective modes, $\hat{\vb*\psi}_{l\to n,1}$, and transfer fields, $\hat{\tau}_{l\to n,1}$, of the three mean production triads, $(f_l,f_n)/f_0=(0,1)$, $(0,2)$, and $(0,3)$. As an aid to interpretability, to remove experimental artifacts present in the data that are exacerbated by the numerical gradient calculations, we apply a low-pass filter to the streamwise wavenumbers of all modes and transfer fields. In figure \ref{fig:turbine-tod_mode}(a), the streamwise velocity components of the recipient and convective modes for $(f_l,f_n)/f_0=(0,1)$ show tip vortex structures that are deformed by the tip shear layer. The vortices experience slow spatial decay, extending past the end of the PIV window at five rotor diameters from the turbine. The transfer field instead has compact support, and is localized mainly in the region $x/D\lesssim3$. The shape of the transfer field is nearly identical to the mean production term computed by \citet{BiswasBuxton2024energy}. While the integral modal transfer of this triad is positive, the spatial modal transfer field reveals that the mean flow donates energy to the rotational frequency, $f_n/f_0=1$, only in the initial shear layer, $x/D\lesssim 1.5$. This is followed by a smaller, inverse contribution for $x/D\gtrsim 1.5$, within which energy is extracted from $f_n/f_0=1$. \citet{BiswasBuxton2024energy} explained the sign change at $x/D\approx 1.5$ in terms of the changing orientation of the vortex triplet shed from the three-bladed rotor, which results in a sign change in the Reynolds stress term, specifically the streamwise and transverse velocity covariance, $\overline{\hat{u}\hat{v}}$. The transfer field, $\hat{\tau}_{l\to n,1}$, which forms a low-rank reconstruction of the production, captures this effect. The streamwise component of the recipient mode, $\hat{\phi}_{n,1}^{(u)}$, optimally represents $\hat u$. The corresponding component of the convective mode, $\hat{\psi}_{l\to n,1}^{(u)}$, optimally represents $\hat u\pdv*{\bar u}{x} + \hat v\pdv*{\bar u}{y}$. A change in the inclination of the recipient mode, relative to the convective mode, therefore alters the sign of the velocity correlations, $\overline{\hat{u}\hat{u}}$ and $\overline{\hat{u}\hat{v}}$, and hence the direction of energy transfer.
Figure \ref{fig:turbine-tod_mode}(a) clearly displays this behavior. For $x/D\lesssim1.5$, the recipient mode is oriented with the convective mode. At $x/D\approx1.5$, the convective mode roughly maintains its angle, while the recipient mode abruptly reorients itself against the convective mode.

The influence of the relative orientation of the modes on energy transfer appears robust at the higher harmonics. For the triad $(f_l,f_n)/f_0=(0,2)$ in figure \ref{fig:turbine-tod_mode}(b), the convective mode once again approximately maintains its inclination throughout the domain. The recipient mode, meanwhile, reorients itself at $x/D\approx2.3$, which coincides with the location where the transfer field switches from negative to positive. For $(f_l,f_n)/f_0=(0,3)$, the modes are more spatially compact, and the effect is subtle. However, the recipient mode appears to undergo orientation change twice, first at $x/D\approx1.7$, and again at $x/D\approx2.3$. Both locations roughly correspond to sign changes in the transfer field.

Figure \ref{fig:turbine-tod_mode}(d--f) shows the modes and transfer fields of the triads $(f_l,f_n)/f_0=(2,3)$, $(3,2)$, and $(3,3)$. The triads $(2,3)$ and $(3,2)$ form a conserved pair in the integral modal energy budget. By way of vortex pairing \citep{BiswasBuxton2024energy}, they transfer energy nonlinearly from $f/f_0=3$ to $f/f_0=2$. The backscatter of energy towards the second harmonic, which is not exogenously forced, explains its prominence in the mode bispectrum in figure \ref{fig:turbine-tod_spec}(a). In general, the transfer fields of a conserved triad pair need not have the same spatial distribution (with the opposite sign): the energy transfer is conserved only in an integral sense; see equation \eqref{pairwise}. In this particular case, however, the transfer fields of $(2,3)$ and $(3,2)$ appear nearly identical in shape. The recipient and convective modes of triad $(2,3)$ both evolve at the frequency $f_n/f_0=3$. As a consequence of the dispersion relation of the tip vortices, which couple the temporal and spatial frequencies of a wave, the recipient and convective modes also share the same streamwise wavenumber. This is also true of the modes of triad $(3,2)$. The recipient and convective modes oscillate at the frequency $f_n/f_0=2$, and have a lower wavenumber than the modes of triad $(2,3)$. For the mean production triads in figure \ref{fig:turbine-tod_mode}(a--c), only the recipient modes are tilted, while the convective modes are roughly aligned with the mean flow. In contrast, for triads $(2,3)$ and $(3,2)$, both recipient and convective modes manifest pronounced tilt.

 \section{Discussion and Summary}\label{sec:discussion}
In this work, we have developed a novel modal decomposition which distills coherent flow structures that optimally account for triadic interactions. For a given triad, triadic orthogonal decomposition decomposes the convective and recipient terms in the momentum equations into a pair of jointly optimal orthogonal modal bases, ranked in terms of the covariance between the projections onto the respective bases. Maximizing the covariance also captures the bispectral statistics that emerge due to prevalent triadic interactions within the flow. By considering a three-wave interaction as the covariance between the convective and recipient terms, we show that the jointly optimal bases are given by the left and right singular vectors of a singular value decomposition of the convective-recipient covariance tensor. The strength of the correlation is given by the singular value, and measures the statistical prevalence of a given triad. The singular values of all triads are presented in the form of a mode bispectrum, providing a succinct indication of active triads. For nonzero recipient frequencies, $f_n\neq0$, the convective and recipient mode pair models the inter-scale momentum transfer as governed by the spectral momentum equations. For $f_n=0$, the mode pair instead contributes to the Reynolds stresses in the RANS equations, thereby acting on the mean flow.

By relating the spectral momentum equations to the spectral kinetic energy equation, we show that a natural consequence of the scale-to-scale modal momentum transfer is the modal energy transfer. The latter is obtained from the singular value and the projection of the convective mode onto the recipient mode. We interpret the spatially-integrated modal energy transfer as a modal energy budget. This interpretation follows from recognizing that the convective term in the kinetic energy equation can be partitioned into spectral TKE contributions from linear advection by the mean, production by the mean, and nonlinear transfer, as well as MKE contributions from mean self-advection, production and nonlinear transfer. The relative importance of each of these budget terms to the overall energy dynamics can be directly read off the bispectrum plane. Specifically, MKE budget terms reside on the $f_n=0$ axis, while spectral TKE budget terms are active off that axis. Within the TKE budget, linear advection by the mean resides on the $f_n=f_l$ diagonal, and production by the mean is found on the $f_l=0$ axis.

As required by the globally conservative nature of the convective term, the nonlinear energy transfer of individual triads, summed over the bispectrum plane, vanishes. Perhaps lesser-known is that triadic energy transfer is conserved also in a pairwise fashion. Since TOD modes form an optimal representation of the recipient and convective fields, under the condition of large separation between the optimal and suboptimal convective-recipient covariances, the integral modal energy transfer is also pairwise conservative. Taking advantage of pairwise conservation, we use the modal energy budget to systematically excavate and quantify the energy donated from an arbitrary frequency, $f_l$, and received by another arbitrary frequency, $f_n$, mediated by a catalyst frequency, $f_k=f_n-f_l$. Positive energy transfers (marked red in this work) in the region $f_n>|f_l|$, or equivalently, negative energy transfers (marked blue) in the region $f_n<|f_l|$, indicate a forward cascade of energy from low to high frequencies. Conversely, negative transfers in $f_n>|f_l|$, or equivalently, positive transfers in $f_n<|f_l|$, signify an inverse cascade of energy from high to low frequencies.

It must be emphasized that finite correlation between the recipient and convective terms does not imply a causal association between the two. That is, the TOD mode bispectrum and modal energy transfer establish a statistical and energetic link between the donor, recipient, and catalyst frequencies, but not a causal link. TOD is therefore not equipped to resolve the questions of which frequency arises first, or which frequency generates which, nor is it obvious that such questions are in fact meaningful. Rather, in this work, a triad interaction is understood as the \emph{collective} action of all three frequency components that contributes to the momentum and energy dynamics of a flow.

Collectively, the scale-to-scale energy exchanges enabled by all donor-recipient pairs are woven into a network of energy transport. Energy enters and leaves the system through the $f_l=0$ axis and the $f_n=f_l$ diagonal. These represent respectively production and linear advection by the mean---linear mechanisms that are unrestrained by energy conservation. Once inside the system, energy is scattered over the nonlinear regions of the bispectrum plane, circulating from donor to recipient and bound by a hierarchy of conservative properties.

We applied TOD to two examples. Notably, in both cases, it was sufficient to focus exclusively on the leading donor-recipient structures identified by TOD. The suboptimal mode bispectra and energy transfer budgets, in contrast to the leading components, exhibited significantly lower amplitudes and lacked any discernible structure.

The first example is DNS data of a laminar unsteady cylinder wake, a benchmark problem for canonical convective flows with well-understood nonlinear dynamics. The leading TOD mode bispectrum reveals that the most dominant triadic interactions occur at integer multiples of the fundamental frequency, confirming that the intrinsic bluff-body vortex-shedding instability mechanism is primarily driven by interactions between the harmonics and their corresponding conjugates. The energy transfer predominantly cascades forward, moving from lower to higher frequencies and thus from larger to smaller spatial scales. This forward energy cascade is evident from the primarily positive energy transfer above the $f_n = f_l$ line and the negative energy transfer below it. Notably, we also observe triads exhibiting the opposite behavior, indicative of an inverse energy transfer.

Each pair of transfer fields associated with a conserved triad pair reveals two distinct structures whose integral sum is zero, demonstrating the balanced contributions of both components. This indicates that, while energy is spatially redistributed through nonlinear interactions within the computational domain, the total energy is conserved. This finding offers deeper insight, emphasizing that a comprehensive analysis of the full structure of the transfer fields, rather than just their integral sums, is essential for fully understanding the physical mechanisms driving these nonlinear interactions.

The second example is TR-PIV data of a turbulent wind turbine wake. This flow is characterized by both deterministic and stochastic components. Moreover, the case is representative of experimental data that are inevitably contaminated by measurement noise. The leading TOD mode bispectrum identifies dominant triads made up of the harmonics of the rotor rotational frequency. Triads with the highest magnitudes are located on the $f_n=0$ axis, indicating a substantial contribution to the mean flow from the Reynolds stress, and on the $f_n=f_l$ diagonal, indicating interaction between the mean flow and the rotor tip vortices. Overall, these linear triadic interactions are significantly stronger than the nonlinear triads in the mode bispectrum. In the turbine wake, the optimal convective-recipient covariance captures the majority of the total covariance. The nonlinear triads within the leading modal energy budget thus successfully recover the pairwise conservative property. The modal budget provides clear evidence of production, linear advection, and nonlinear transfer between the harmonic frequencies. Production by the mean injects energy into the rotational frequency, but extracts energy from its higher harmonics. Linear advection by the mean injects energy into all harmonics. Negative and positive nonlinear transfers above and below the $f_n=f_l$ diagonal, respectively, exemplify an inverse energy cascade. The recipient modes reveal vortex shedding from the rotor tips. Like the cylinder wake, the modal transfer fields have compact support. The streamwise locations of sign changes in the tranfer fields are explained in terms of re-orientation of the recipient mode against the convective mode, or vice versa. The transfer field corresponding to the inverse transfer from the third to the second harmonic and the transfer field of the conserved pair appear nearly identical in spatial structure. As is almost always the case with experimental diagnostics, some information is lost when two-dimensional data are captured from a statistically three-dimensional flow, such as the turbine wake. This inherently limits the terms in the momentum equations that can be modeled using TOD. In particular, out-of-plane velocity and velocity gradients, which contribute to momentum and energy transfer, cannot be captured. That the method nevertheless succeeds in uncovering the nonlinear dynamics of the turbine wake supports its broad applicability to data sets in which the full flow state is only partially observable.

To extract maximal performance from TOD, the following best practices should be adhered to. TOD, much like other frequency-domain methods, benefits greatly from the careful selection of spectral estimation parameters that strikes a good balance between minimizing statistical variance and bias. The same recommendations that apply to the standard SPOD algorithm \citep{towne2018spectral,schmidt2020guide} based on Welch's overlapped segment averaging (WOSA; \citep{welch1967use}) should be followed here. For tonal or broadband-tonal flows characterized by some fundamental frequency or period, the need to resolve harmonic peaks accurately places additional constraints on the segment length, $N_f\Delta t$. To center the peaks in the frequency bins of the DFT and thus avoid spectral leakage, this length should be (close to) an integer multiple of the fundamental period. This renders the data (nearly) periodic within each segment. For periodic data, a rectangular window should be used. More detail on spectral estimation for data with time-periodic statistics can be found in \citet{HeidtColonius2024JFM}. For data with particularly challenging demands on spectral resolution and statistical convergence, TOD may benefit from replacing the standard Welch estimator with a multitaper estimator \citep{Thomson1982IEEE,RiedelSidorenko1995IEEE} by analogy with multitaper SPOD \citep{Schmidt2022TCFD,YeungSchmidt2024TCFD}. Other techniques that aim to improve the convergence of SPOD may also be applicable to TOD, including recent algorithms proposed by \citet{BlancoEtAl2022JFM}, \citet{colanera2024robust}, \citet{heidt2024optimal}, and \citet{YeungSchmidt2024TCFD}.

If primitive variables are not available, as is common with experimental data such as schlieren images, it will not be possible to compute the terms in the momentum equations. For these data sets, we can still apply TOD by forgoing its specialization to the momentum and kinetic energy equations. For a general observable, $\vb*q(\vb*x,t)$, we can redefine the cross-bispectral covariance tensor as
\begin{align}\label{eq:qHadamardq}
    \vb*S(\vb*x,\vb*x') = \mathrm{E}\qty{\vb*{\hat q}_{(n-l)\circ l}(\vb*x)\vb*{\hat q}^{\mathrm{H}}_{n}(\vb*x')} \qq{for each} (l,n),
\end{align}
where $\vb*{\hat q}_{(n-l)\circ l} \equiv \vb*{\hat q}_{n-l}\circ \vb*{\hat q}_l$, and $\circ$ denotes the Hadamard product. The treatment of nonlinear coupling in equation \eqref{eq:qHadamardq} is closely connected to BMD \citep{schmidt2020bispectral}, which also uses the term $\vb*{\hat q}_{(n-l)\circ l}$. The SVE of $\vb*S$ yields a mode bispectrum and two sets of modal bases, with the same properties as the algorithm in \S \ref{sec:method}. Without access to velocity, the modal energy flow analysis in \S \ref{sec:modalEnergyFlow} is not possible. Nevertheless, even in this restricted setting, we have confirmed that TOD succeeds in identifying dominant triads, as triad interactions give rise to correlation among frequency components of the observable, which TOD detects.

Extensions of TOD to study other types of dynamics may be envisioned. By construction, TOD expansion efficients are uncorrelated; see equation \eqref{eq:coeffsUncorrelated}. As we saw in \S \ref{subsec:directivity}, a direct consequence is that, on average, no energy is exchanged between leading and suboptimal recipient and convective modes. In turbulent flows with a time-periodic mean flow, each leading TOD mode may be interpreted as a deterministic component of the mean, and the suboptimal modes as the stochastic turbulence. The uncorrelatedness between leading and suboptimal modes thus translates to the uncorrelatedness between the deterministic and stochastic components of the flow. In contrast, the findings of \citet{HeidtColonius2024JFM} in relation to CS-SPOD formalizes the concept that finite correlation can arise between a periodic mean and the underlying turbulent statistics. Momentum and energy transfers between deterministic motions and the turbulent fluctuations about them are well-trodden territory, and are typically considered by subjecting the equations of motion to a triple decomposition \citep{HussainReynolds1970JFM}, producing coupled equations for the deterministic (organized) and stochastic (turbulent) components \citep{ReynoldsHussain1972JFM}. Recent applications include \citet{OberleithnerEtAl2014JFM} and \citet{heidt2023cyclostationary} for turbulent jets, and \citet{KinjangiFoti2023JFM} for a turbulent cylinder wake. For the examples in this work, the large separation between the leading and suboptimal convective-recipient covariances in the TOD mode bispectrum demonstrates that the dynamics of each flow is already well-described by only the organized motion. In flows where this is not the case, we can easily investigate organized motion-turbulence interactions using TOD by considering the momentum equations for the turbulent component \citep[equation 2.8]{ReynoldsHussain1972JFM}, in place of the full momentum equations \eqref{mom_eqn}. This may be accomplished by removing the periodic mean from the recipient frequency and either the donor or the catalyst frequency of each triad. For example, removing the periodic mean from the recipient and catalyst, while keeping it for the donor, extracts the correlation and transfer between a deterministic-stochastic donor-recipient pair.

In addition to the Navier-Stokes equations, TOD can be generalized to other triadically nonlinear phenomena. For instance, it is straightforward to incorporate velocity and magnetic fields into the recipient and convective terms, then apply TOD to study spectral kinetic and magnetic energy transfers in magnetohydrodynamic (MHD) turbulence.

TOD also opens the door to novel reduced-order models (ROM) for which the modal bases, rather than being constructed in a linear framework, are instead tailored to triadic interactions and thus optimal in capturing the convective nonlinearity of the Navier-Stokes equations. As suggested by \citet{schmidhenningson2001springer}, the dynamical equations of a ROM should preserve the conservative property of the nonlinear term. They showed that the nonlinear part of truncated equations that use conserved sets of triads---like the ones in \S \ref{subsec:conservation}---as building blocks, indeed conserves energy. For data with a large separation between the leading and suboptimal TOD singular values, these conservative principles are also satisfied by the leading TOD modes and transfer fields. TOD should therefore facilitate the construction of energy-conserving ROMs. When selecting a truncated set of triads to include in a model, it is crucial to consult the mode bispectrum in addition to the modal energy budget. While the modal budget provides copious physical insights, it is the spatial integral of the transfer field and thus prone to cancellation. A triad with small integral energy transfer can have a large singular value (see figures \ref{fig:Cylinder-tod_spec} and \ref{fig:turbine-tod_spec}). In contrast, singular values in the mode bispectrum, which measure covariance, directly convey the statistical significance of each triad.


 \setcounter{section}{0}
\setcounter{figure}{0}
\setcounter{equation}{0}
\renewcommand{\theequation}{{\rm A}.\arabic{equation}}

\section*{Acknowledgments}
OTS acknowledges support from AFOSR grant FA9550-22-1-0541 and NSF grant CBET 2046311. BY gratefully acknowledges support from ONR grant N00014-23-1-2457. The authors are grateful to Neelakash Biswas and Oliver Buxton at Imperial College London for generously sharing their wind turbine PIV data. OTS also extends thanks to Julian Domaradzki for insightful discussions on triadic interactions.


\bibliography{main}

\end{document}